\documentclass[12pt,a4paper]{article}
\pdfoutput=1

\usepackage[utf8]{inputenc}
\usepackage[T1]{fontenc}

\usepackage{cite}
\usepackage{amsmath}
\usepackage{color}
\usepackage{amsfonts}
\usepackage{amssymb}
\usepackage{graphicx}
\usepackage{geometry}
\usepackage{amssymb,epsfig,subfigure}
\usepackage{hyperref}
\usepackage{url}
\usepackage{comment}
\usepackage[font=footnotesize]{caption}
\usepackage{slashed}

\makeatletter
\renewcommand\section{\@startsection {section}{1}{\z@}%
                                 {-3.5ex \@plus -1ex \@minus -.2ex}
                                   {2.3ex \@plus.2ex}%
                                   {\normalfont\large\bfseries}}
\renewcommand\subsection{\@startsection{subsection}{2}{\z@}%
                                   {-3.25ex\@plus -1ex \@minus -.2ex}%
                                     {1.5ex \@plus .2ex}%
                                     {\normalfont\bfseries}}
\renewcommand\subsubsection{\@startsection{subsubsection}{3}{\z@}%
                                   {-3.25ex\@plus -1ex \@minus -.2ex}%
                                     {1.5ex \@plus .2ex}%
                                     {\normalfont\itshape}}
\makeatother

\def\pplogo{\vbox{\kern-\headheight\kern -29pt
\halign{##&##\hfil\cr&{\ppnumber}\cr\rule{0pt}{2.5ex}&\ppdate\cr}}}
\makeatletter
\def\ps@firstpage{\ps@empty \def\@oddhead{\hss\pplogo}%
  \let\@evenhead\@oddhead 
}
\def\maketitle{\par
 \begingroup
 \def\thefootnote{\fnsymbol{footnote}}
 \def\@makefnmark{\hbox{$^{\@thefnmark}$\hss}}
 \if@twocolumn
 \twocolumn[\@maketitle]
 \else \newpage
 \global\@topnum\z@ \@maketitle \fi\thispagestyle{firstpage}\@thanks
 \endgroup
 \setcounter{footnote}{0}
 \let\maketitle\relax
 \let\@maketitle\relax
 \gdef\@thanks{}\gdef\@author{}\gdef\@title{}\let\thanks\relax}
\makeatother

\numberwithin{equation}{section}

\newcommand\eea{\end{eqnarray}}
\newcommand\bea{\begin{eqnarray}}

\def\beq{\begin{equation}}
\def\eeq{\end{equation}}

\newcommand{\be}{\begin{equation}}
\newcommand{\ee}{\end{equation}}
\newcommand{\ba}{\begin{align}}
\newcommand{\ea}{\end{align}}
\newcommand{\bg}{\begin{gather}}
\newcommand{\eg}{\end{gather}}
\newcommand{\bseq}{\begin{subequations}}
\newcommand{\eseq}{\end{subequations}}

\newcommand{\Tr}{{\rm Tr}}
\renewcommand{\Im}{\mathop{\rm Im}\nolimits}
\renewcommand{\Re}{\mathop{\rm Re}\nolimits}
\renewcommand{\t}{\tilde}

\newcommand{\tr}{{\rm tr}}

\newcommand{\sgn}{\text{sgn}}
\newcommand{\bp}{\textbf{p}}
\newcommand{\bx}{\textbf{x}}
\newcommand{\by}{\textbf{y}}

\begin{document}
\setcounter{page}0
\def\ppnumber{\vbox{\baselineskip14pt
}}
\def\ppdate{
} \date{}

\author{Lucas Daguerre$^1$, Raimel Medina$^2$, Mario Sol\'is$^{3,1}$, Gonzalo Torroba$^{3,1}$\\
[7mm] \\
{\normalsize \it $^1$ Instituto Balseiro, UNCuyo and CNEA}\\
{\normalsize \it S.C. de Bariloche, R\'io Negro, R8402AGP, Argentina}\\
{\normalsize \it $^2$IST Austria }\\
{\normalsize \it Am Campus 1, 3400 Klosterneuburg, Austria}\\
{\normalsize \it $^3$Centro At\'omico Bariloche and CONICET}\\
{\normalsize \it S.C. de Bariloche, R\'io Negro, R8402AGP, Argentina}\\
}

\bigskip
\title{\bf  Aspects of quantum information in \\ finite density field theory
\vskip 0.5cm}
\maketitle


\begin{abstract}
We study different aspects of quantum field theory at finite density using methods from quantum information theory. For simplicity we focus on massive Dirac fermions with nonzero chemical potential, and work in $1+1$ space-time dimensions. Using the entanglement entropy on an interval, we construct an entropic $c$-function that is finite. Unlike what happens in Lorentz-invariant theories, this $c$-function exhibits a strong violation of monotonicity; it also encodes the creation of long-range entanglement from the Fermi surface. Motivated by previous works on lattice models, we next calculate numerically the Renyi entropies and find Friedel-type oscillations; these are understood in terms of a defect operator product expansion. Furthermore, we consider the mutual information as a measure of correlation functions between different regions. Using a long-distance expansion previously developed by Cardy, we argue that the mutual information detects Fermi surface correlations already at leading order in the expansion. We also analyze the relative entropy and its Renyi generalizations in order to distinguish states with different charge and/or mass. In particular, we show that states in different superselection sectors give rise to a super-extensive behavior in the relative entropy. Finally, we discuss possible extensions to interacting theories, and argue for the relevance of some of these measures for probing non-Fermi liquids.
\end{abstract}
\bigskip

\newpage

\tableofcontents

\vskip 1cm

\section{Introduction}\label{sec:intro}

The last decades have seen groundbreaking progress in nonperturbative aspects of quantum field theory (QFT) using methods from quantum information theory. This started with Srednicki's calculation of entanglement entropy for a free scalar field, exhibiting the area law~\cite{Srednicki:1993im}. Measures based on the entanglement entropy together with unitarity and causality have established the irreversibility of the renormalization group in $d=1+1,\,2+1$ and $3+1$ spacetime dimensions for relativistic QFT~\cite{Casini:2004bw, Casini:2012ei, Casini:2017vbe}.\footnote{The $C$ and $A$ theorems for $d=1+1$ and $d=3+1$ respectively were proved originally using local correlators in \cite{Zamolodchikov:1986gt} and \cite{Komargodski:2011vj}, respectively.} Results from conformal field theory (CFT)~\cite{Calabrese:2009qy} as well as free field theory~\cite{Casini:2009sr} have led to new insights into the structure of quantum field theory. Properties of relative entropy have led to energy bounds in QFT, such as the Bekenstein bound~\cite{Casini:2008cr}, the generalized second law~\cite{Wall:2011hj}, the ANEC~\cite{Faulkner:2016mzt} and the QNEC~\cite{Balakrishnan:2017bjg}. The observation of the area law in local models has also led to powerful numerical methods for finding ground-states~\cite{Orus:2013kga}. And starting from the proposal of Ryu and Takayanagi~\cite{Ryu:2006bv, Ryu:2006ef} that entanglement entropy in a CFT can be computed from a surface of minimal area in anti-de-Sitter spacetime, ideas from quantum information have led to fundamental developments in holography and quantum gravity.

Most of these results have been so far restricted to relativistic QFT. The reason is that constraints from Lorentz symmetry and causality play a key role in the above approaches. In contrast, much less is known about nonrelativistic QFT and, in particular, in field theory at finite density as would be relevant for describing the continuum limit of quantum matter. It has been suggested that entanglement entropy can present a non-monotonic behavior in certain nonrelativistic models and that this would imply that the renormalization group may not be generically irreversible beyond Lorentz-invariant theories~\cite{Swingle:2013zla}. Works on entanglement entropy for relativistic fermions with finite charge include~\cite{Ogawa:2011bz, Belin:2013uta, Herzog:2013py, cardyconf, Kim:2017xoh, Kim:2017ghc}. Some results on theories with nontrivial dynamical exponents include~\cite{Fradkin:2009dus, Hsu:2008af, He:2017wla, MohammadiMozaffar:2017nri}.

The goal of this work is to determine what quantum information can tell us about relativistic field theory at finite density, obtained by turning on a chemical potential. In some sense this is in between a relativistic field theory and completely nonrelativistic models: the chemical potential is a relevant perturbation and hence at very short distances we expect to recover Poincare invariance. It is therefore a natural starting point to try to extend results from the relativistic context. Such models can also have Fermi surfaces, and their study is then important for experimental applications (we will in fact derive new predictions in this direction).

A key lesson from Lorentz invariant analyses is that, from the point of view of quantum information measures, free QFTs provide a nontrivial playground for obtaining results that usually apply more broadly. This point of view has been reviewed and advocated in~\cite{Casini:2009sr}. Motivated by this, in this paper we study free Dirac fermions at finite density, employing both analytical methods in quantum information theory as well as numerical simulations. 
Furthermore, we restrict to the simplest case of $d=1+1$ space-time dimensions. The reason for this is that numerical calculations are easiest in this case; in spite of these simplifications, the results are already nontrivial. We plan to extend our analysis to higher dimensions in future work. We will find that information-theoretic measures exhibit a behavior that is quite different from that in relativistic theories; this is related to the fact that Fermi surfaces give rise to low energy excitations with long range entanglement.

After reviewing basic properties of Dirac fermions at finite density in Sec.~\ref{sec:Dirac}, the rest of the work will be devoted to a detailed analysis of different quantum information measures and their implications in QFT. The dynamics is determined by two relevant parameters, the mass $m$ and the charge that is proportional to the Fermi momentum $k_F$.
We start with the entanglement entropy in Sec.~\ref{sec:EE}. We construct a cutoff independent entropic $c$-function, and prove that it is non-monotonic. It reveals the competition between $m$ and $k_F$ along the renormalization group flow, and encodes in an explicit way the creation of entanglement and long-range correlations associated to the Fermi surface. In Sec.~\ref{sec:REE} we study the Renyi entropies at finite density. Previous works on lattice models, such as~\cite{Calabrese2010ParityEI}, have found oscillatory behavior in these measures, and this was related to Friedel oscillations in~\cite{Swingle_2013}. We analyze when and how these arise in continuum QFT, and explain their origin in terms of a defect-based operator product expansion. Next, in Sec.~\ref{sec:mutual} we consider the mutual information (and its Renyi generalization) between two intervals. We follow~\cite{Cardy.esferaslejanas} and represent each interval in terms of an operator product expansion with local operators.  We prove that the mutual information detects already at leading order the presence of a Fermi surface, and we suggest why this is an interesting measure to probe non-Fermi liquids. The analytic results are also verified numerically. Finally, in Sec.~\ref{sec:relrenyi} we study the relative entropy, and a certain one-parameter generalization of it that interpolates to the fidelity. These quantities measure the distinguishability between states. We apply them to study the distinguishability between states in two different cases: density matrices in distinct superselection charge sectors, and density matrices in the same charge sector but with different relevant deformations. We also compare with numerical results. We end in Sec.~\ref{sec:concl} with a summary and discussion of future directions; some properties that we require about lattice models are presented in Appendix \ref{app:lattice}.

\section{Free Dirac fermions at finite density}\label{sec:Dirac}

In this section we briefly review the physics of Dirac fermions at finite density. The rest of the work is devoted to analyzing measures from quantum information in this theory.

\subsection{Continuum theory}\label{subsec:cont}

A free Dirac fermion has a $U(1)$ symmetry $\psi \to e^{i \theta} \psi$. A finite charge density $n_e= \langle \psi_{\alpha}^\dag \psi_{\alpha} \rangle$ is obtained by turning on a chemical potential $\mu_F$, which appears as an expectation value for a background gauge field with nonzero time component.  The action (with signature $g_{\mu \nu}=(+-\ldots -)$) reads
\begin{equation}
S = \int dt \,d^{d-1} x\,\left(\bar{\psi}(i\slashed{\partial}-m)\psi +\mu_F \psi^{\dag}\psi \right)\,.
\label{eq:Sdirac}
\end{equation}
Both  the mass and chemical potential are relevant operators that induce a nontrivial renormalization group (RG) flow from the UV CFT of a massless Dirac fermion. We follow the conventions in~\cite{Peskin:1995ev}, and our choice of Dirac matrices is described in Appendix~\ref{app:lattice}.

The energy eigenvalues are
\be\label{eq:Epm}
E_\pm = - \mu_F \pm \sqrt{\bp^2 +m^2}\,.
\ee
In what follows we choose $\mu_F>0$. Then $E_-$ is always negative and as usual gives rise to antiparticles; the branch $E_+$ has vanishing energy at a finite momentum $|\bp|= k_F$, with
\be\label{eq:kF}
k_F = \sqrt{\mu_F^2 - m^2}\,.
\ee
This defines a spherical Fermi surface. The ground state, described in more detail in Sec.~\ref{subsec:latt_mod}, is obtained by filling these states up to $E_+=0$. Note that we need
\be
|\mu_F| > |m|
\ee
otherwise the Fermi level is below the gap, and no charge is added to the system.

In order to compute the time-ordered correlator, the appropriate $i \epsilon$ prescription moves the poles for the Green's function to
\be
p_0= E_\pm - i \epsilon\,\sgn(E_\pm)\,.
\ee
In other words~\cite{abrikosov2012methods, Peskin:1995ev}, 
\be
\frac{1}{\t p^2 - m^2} \,\to\, \frac{1}{2 \sqrt{\bp^2 +m^2}} \left(\frac{1}{p^0- E_++i \epsilon\, \sgn E_+}-\frac{1}{p^0- E_-+i \epsilon\, \sgn E_-} \right)\,,
\ee
for the residue calculation, and we have defined $\tilde{p}_{\mu}=(p_0+\mu_F,p_i)$.
The correlator then reads
\begin{equation}
\langle T \bar \psi_\alpha(x)  \psi_\beta(y) \rangle=-\int \frac{d^d p}{(2\pi)^d} \frac{i(\slashed{\tilde{p}}+m)_{\beta \alpha}}{2 \sqrt{\bp^2 +m^2}} e^{ip(x-y)}\left(\frac{1}{p^0- E_++i \epsilon\, \sgn E_+}-\frac{1}{p^0- E_-+i \epsilon\, \sgn E_-} \right)\,.
\end{equation}

For entanglement calculations below, let us compute the equal-time correlator. We can choose 
$x^0 - y^0 \to 0^+$ so that $\langle T \bar \psi(x)  \psi(y) \rangle= \langle \bar \psi(\bx) \psi(\by) \rangle$. Closing the $p^0$ contour in the upper half plane, where the Fourier factor $e^{i p^0(x^0-y^0)}$ gives an exponential suppression, obtains
\bea\label{eq:Gf}
\langle \bar \psi(\bx) \psi(\by) \rangle&=&  \int \frac{d^{d-1}p}{(2\pi)^{d-1}}e^{-i\textbf{p}\cdot(\textbf{x}-\textbf{y})} \frac{1}{2 \sqrt{\bp^2 +m^2}}  \Bigg \lbrace \Theta(-E_+) \left( \sqrt{\bp^2 +m^2} \gamma^0+ \gamma^i p^i+m \right)\nonumber\\
&&\qquad - \left(-\sqrt{\bp^2 +m^2} \gamma^0+ \gamma^i p^i+m \right) \Bigg \rbrace\,.
\eea
In order to compute the reduced density matrix, it will actually be more convenient to work in terms of the correlator
\bea\label{eq:C1}
 C(\bx-\by)= \langle \psi^\dag(\bx) \psi(\by) \rangle&=&\int \frac{d^{d-1}p}{(2\pi)^{d-1}}e^{-i\textbf{p}\cdot(\textbf{x}-\textbf{y})}  \Bigg \lbrace \Theta(-E_+) \left(\frac{1}{2}+
 \frac{ \gamma^0\gamma^i p^i+\gamma^0 m}{2 \sqrt{\bp^2 +m^2}}  \right)\nonumber\\
&&\qquad  + \left(\frac{1}{2}-
 \frac{ \gamma^0\gamma^i p^i+\gamma^0 m}{2 \sqrt{\bp^2 +m^2}}  \right) \Bigg \rbrace\,.
\eea

We recognize the finite density correlator as a deformation of the relativistic result,
\be\label{eq:CkF}
 C_{k_F}(\bx-\by)= C_0(\bx-\by) +\int_{|\bp|< k_F} \frac{d^{d-1}p}{(2\pi)^{d-1}}e^{-i\textbf{p}\cdot(\textbf{x}-\textbf{y})}  \left(\frac{1}{2}+
 \frac{ \gamma^0\gamma^i p^i+\gamma^0 m}{2 \sqrt{\bp^2 +m^2}}  \right)\,,
\ee
with
\be
 C_0(\bx-\by)=  \int \frac{d^{d-1}p}{(2\pi)^{d-1}}e^{-i\textbf{p}\cdot(\textbf{x}-\textbf{y})} \left(\frac{1}{2}-
 \frac{ \gamma^0\gamma^i p^i+\gamma^0 m}{2 \sqrt{\bp^2 +m^2}}  \right)\,.
\ee

\subsection{Lattice model}\label{subsec:latt_mod}

The entanglement and Renyi entropies diverge in continuum QFT. A physical way to regulate them is to put the theory on a lattice, and we will then use the real time method~\cite{peschel2003,  Casini:2009sr} to evaluate the entanglement measures. In what follows we restrict to $d=1+1$ space-time dimensions.

We work with an infinite spatial lattice, $x^1 = n\,a$, $n \in \mathbb Z$, and set the lattice spacing $a=1$. Symmetrizing the spatial derivative in (\ref{eq:Sdirac}) and discretizing the derivatives as $\partial_1 \psi(x) \to (\psi_{n+1}- \psi_n)/a$, the lattice Hamiltonian reads
\begin{equation}\label{eq:Hlattice}
\mathcal{H}=\sum_n\left(-\frac{i}{2}(\psi_n^{\dag}\gamma^0\gamma^1(\psi_{n+1}-\psi_n)-\text{h.c.})+m\psi_n^{\dag}\gamma^0\psi_n-\mu_F \psi_n^{\dag}\psi_n\right)=\sum_{i,j} \psi_i^{\dag} H_{ij}  \psi_j,
\end{equation}
where $\lbrace(\psi_i)_{\alpha},(\psi^{\dag}_j)_{\beta}\rbrace=\delta_{ij}\delta_{\alpha \beta}$, being $\alpha$, $\beta$ spinor indices. We will now determine the ground state and  the equal time fermion Green's function.

In momentum space,
\begin{equation}
\psi_n=\int_{-\pi}^\pi \frac{dk}{2\pi}\, \varphi_k e^{ikn}\,,
\end{equation}
the Hamiltonian becomes
\begin{equation}
\mathcal{H}=\int_{-\pi}^\pi  \frac{dk}{2 \pi}\,\varphi_k^{\dag}\left(\sin(k)\gamma^0\gamma^1+m\gamma^0-\mu_F \right)\varphi_k=\int \frac{dk}{2 \pi}\: \varphi_k^{\dag}M(k)\varphi_k.
\end{equation}
The energy eigenvalues are given by
\be\label{eq:eppm}
\epsilon_{\pm}(k)=-\mu_F \pm \sqrt{\sin(k)^2+m^2}\,,
\ee
and the corresponding normalized eigenvectors are denoted by $v_{\pm}(k)$ respectively. The Hamiltonian is then diagonalized in the new basis
\begin{equation}\label{eq:cphi}
\begin{pmatrix}c_{k,+}  \\ c_{-k,-}^\dag \end{pmatrix}=U^{\dag}(k)\varphi_k\;,
\end{equation}
with
\be
U(k)=\left(v_+(k) \;,\; v_-(k)\right)
\ee
the unitary matrix of eigenvectors of $M(k)$, and becomes
\begin{equation}
\mathcal{H}=\int_{-\pi}^\pi  \frac{dk}{2 \pi}\: \left(\epsilon_+(k)c^{\dag}_{k,+}c_{k,+}+(-\epsilon_-(k))c^{\dag}_{k,-}c_{k,-}\right)
\end{equation}
after subtracting the zero-point energy.
Unitarity of the transformation guarantees that $\{c_{k,a},\:c^{\dag}_{p,b} \}=2\pi\delta(k-p)\delta_{ab}$, $a,\,b=\pm$. 

When $\mu_F=0$, the vacuum state is the zero-particle state $|0\rangle$, annihilated by all $c_{k, \pm}$. Once $\mu_F>0$, the new ground state is obtained by filling the negative energies $\epsilon_+(k)<0$ in the particle band:
\begin{equation}\label{eq:Gdef}
|G\rangle =\prod_{k\;,\;\epsilon_+(k)\leq 0}c^{\dag}_{k,+}|0\rangle.
\end{equation}
The particles then fill a Fermi surface with Fermi momentum
\be\label{eq:kFlattice}
k_F= \arcsin \left( \sqrt{\mu_F^2-m^2}\right)\,,
\ee
and another Fermi surface at $ \pm (\pi-k_F)$. 

The continuum limit is obtained by sending all the (dimensionless) energy quantities in (\ref{eq:eppm}) to zero, keeping their ratios $k/m$ and $\mu_F/m$ fixed. This then recovers (\ref{eq:Epm}).\footnote{Equivalently, we can reintroduce the lattice spacing, to obtain the dispersion relation $\epsilon_\pm(k) = - \mu_F + \sqrt{\frac{\sin(k a)^2}{a^2}+m^2}$. Taking $a \to 0$ gives (\ref{eq:Epm}).} However, other properties are specific to the lattice. One is the well-known fermion doubling: due to the periodicity of the dispersion relation with the lattice momentum, the model gives two Fermi surfaces; see Appendix \ref{app:lattice} for more details.
 Another property is that once $\mu_F= \sqrt{1+m^2}$, the particle band is completely filled; so in our calculations below we will always take $\mu_F \le \sqrt{1+m^2}$. This restriction disappears in the continuum limit, because the momentum cutoff $|k | < \pi/a \to \infty$.

We are now ready to compute the correlation function in the lattice model,
\begin{equation}
C_{ij}=\langle G|\psi^{\dag}_i \psi_j|G\rangle \equiv \langle \psi^{\dag}_i \psi_j\rangle=\int \frac{dk}{2 \pi}\:e^{-ik(i-j)}
 \langle \varphi^{\dag}(k) \varphi(k)\rangle.
\end{equation}
Using the anticommutation relations and noticing the negative energy states filled in (\ref{eq:Gdef}), we have
\begin{equation}
\langle \varphi^{\dag}(k) \varphi(k)\rangle=v^{\dag}_{+}(k)v_+(k)\Theta(-\epsilon_+(k))+v^{\dag}_{-}(k)v_-(k)\Theta(-\epsilon_-(k))\,.
\label{eq:corr_func_mom}
\end{equation}
These terms are computed in (\ref{eq:vp}) and (\ref{eq:vm}) in Appendix \ref{app:lattice}, with the result
\begin{align}\label{eq:Cij}
C_{ij}&=\int_{-\pi}^{\pi} \frac{dk}{2 \pi}\,e^{-ik(i-j)}\, \left(\frac{1}{2}\mathbb{I}-\frac{\sin(k)\gamma^0\gamma^1+m\gamma^0}{2\sqrt{m^2+\sin(k)^2}}\right) \nonumber \\
&+\int_{-\pi}^{\pi} \frac{dk}{2 \pi}\,e^{-ik(i-j)}\,\Theta(-\epsilon_+(k)) \left(\frac{1}{2}\mathbb{I}+\frac{\sin(k)\gamma^0\gamma^1+m\gamma^0}{2\sqrt{m^2+\sin(k)^2}}\right)\,.
\end{align}
The first line is independent of the chemical potential, since $\Theta(-\epsilon_-(k))=1$ for all $k$ in (\ref{eq:corr_func_mom}). This is the contribution from the filled antiparticle band. The second line is the part that encodes the contribution from the Fermi surface of particles.
This is the discrete version of (\ref{eq:C1}), and gives the right continuum limit.

\subsection{Fermion dynamics}\label{subsec:dynamics}

Since the model is Gaussian, the dynamics is completely determined by the propagator (\ref{eq:C1}) or its lattice version (\ref{eq:Cij}). We will now analyze the new features brought in by finite density, and the interplay between the two relevant scales $m$ and $k_F$. 

First, let us consider the massless limit $m/k_F \to 0$. As described in Appendix~\ref{app:lattice}, it is useful to choose the chiral basis $\gamma^0=\sigma^1$, $\gamma^1=i\sigma^2$. Then $\gamma^3=\gamma^0\gamma^1=-\sigma^3$. In terms of left and right movers, $\psi= (\psi_L, \psi_R)$, we obtain
\be
S= \int dx^0 dx^1\left(\psi_L^\dag (i(\partial_0-\partial_1) + \mu_F)\psi_L +\psi_R^\dag (i(\partial_0+\partial_1) + \mu_F)\psi_R \right)\,.
\ee
The chiral fermions decouple; each one has a semi-infinite Fermi surface that ends at $|p^1|= \mu_F$. Furthermore, in this case the Fermi momentum is simply $k_F = \mu_F$.

The low energy theory is obtained by redefining 
\be\label{eq:pperp}
p^1= \pm k_F + p_\perp\;,
\ee 
and restricting to $|p_\perp| \ll |k_F|$. This gives two chiral fermions moving at the speed of light,
\be\label{eq:eff-chiral}
S= \int \frac{dp^0 dp_\perp}{(2\pi)^2}\left(\psi_L^\dag (p^0+p_\perp)\psi_L +\psi_R^\dag (p^0-p_\perp)\psi_R \right)\,.
\ee
There is an equivalent way to think about this, which will be important below. Due to the additional chiral symmetry, the theory has now two $U(1)_L \times U(1)_R$ symmetries, that rotate $\psi_{L, R}$ independently. The chemical potential can be removed by a unitary local transformation,
\be\label{eq:localU}
\psi_L(x^1) = e^{-i k_F x^1}\psi_L'(x^1)\;,\;\psi_R(x^1) = e^{i k_F x^1}\psi_R'(x^1)\,.
\ee
This is not a symmetry, because the phase rotation depends on $x^1$; it maps the theory with chemical potential to one with no chemical potential. This is equivalent to changing the origin of momentum space as in (\ref{eq:pperp}).

The equal time propagators for chiral fermions at zero density are
\be
\langle {\psi'}^{\dag}_L(x^1) \psi_L'(y^1) \rangle = -\frac{i}{2\pi(x^1-y^1)}\;,\;\langle {\psi'}^\dag_R(x^1) \psi_R'(y^1) \rangle =\frac{i}{2\pi(x^1-y^1)}\,.
\ee
We can obtain the corresponding correlators at finite density by applying (\ref{eq:localU}) to this result:
\be
\langle \psi^\dag_L(x^1) \psi_L(y^1) \rangle = -\frac{i}{2\pi(x^1-y^1)}e^{i k_F (x^1-y^1)}\;,\;\langle \psi^\dag_R(x^1) \psi_R(y^1) \rangle = \frac{i}{2\pi(x^1-y^1)}e^{-i k_F (x^1-y^1)}\,.
\ee
The same result can be obtained directly from (\ref{eq:C1}).
This shows terms that oscillate with frequency $k_F$ -- a consequence of the Fermi surface. This is a key difference with the relativistic theory, and similar oscillating terms will be found also away from the massless limit.

Let us now consider the opposite nonrelativistic limit, $m/k_F \to \infty$. At energies and momenta much smaller than the mass, the dispersion relation (\ref{eq:Epm}) becomes
\be
E_\pm \approx -\mu_F \pm |m| \pm \frac{p^2}{2 |m|}\,.
\ee
We choose the mass and chemical potential to be positive. Defining 
\be\label{eq:tmuF}
\mu_F = m + \tilde \mu_F
\ee 
gives
\be
E_+ \approx \tilde \mu_F + \frac{p^2}{2 m}\;,\;-E_- \approx 2m + \tilde \mu_F +\frac{p^2}{2 m}\,.
\ee
The antiparticles with energy $-E_- \approx 2m$ decouple from the low energy theory, while the particles reproduce the dispersion relation of a nonrelativistic fermion, with $\tilde \mu_F$ playing the role of the chemical potential in the nonrelativistic theory. This fermion is spinless, and we will denote it by $\t \psi$. It is related to the Dirac fermion by diagonalizing the Hamiltonian as in (\ref{eq:cphi}) and taking $m \to \infty$; this gives
\be
\t \psi(x) = \frac{1}{\sqrt{2}} (\psi_L(x) + \psi_R(x))\,.
\ee
The low energy effective theory is
\be\label{eq:nonrel2}
S_\text{eff}= \int dx^0 dx^1\, \t \psi^\dag \left( i \partial_0 + \frac{\partial_1^2}{2m} -\t \mu_F\right) \t \psi\,.
\ee
It describes a Fermi surface (two points) with Fermi momentum
\be
\frac{k_F^2}{2m}= \t \mu_F\,.
\ee
Linearizing the dispersion relation around each of the Fermi points gives
\be
\frac{p^2}{2m} - \t \mu_F = \pm v_F p_\perp\;,\;p = \pm (k_F + p_\perp)\;,\;v_F= \frac{k_F}{m}\,.
\ee
Restricting to momenta $p_\perp$ much smaller than $k_F$, and denoting the fermion near the left or right Fermi points by $\t \psi_{L,R}$ respectively, obtains
\be\label{eq:eff-nonrel}
S= \int \frac{dp^0 dp_\perp}{(2\pi)^2}\left(\t \psi_L^\dag (p^0+v_F p_\perp)\t \psi_L +\t \psi_R^\dag (p^0-v_F p_\perp)\t \psi_R \right)\,.
\ee
This is the same as the theory of two chiral fermions (\ref{eq:eff-chiral}), except that now they move at the Fermi velocity $v_F$ instead of the speed of light. The massive theory at zero density becomes trivially gapped at long distances, but at finite density $k_F>0$, two massless chiral fermions emerge. This will lead to strong signatures in quantum information measures.

\section{Entanglement entropy}\label{sec:EE}

We now begin our analysis of quantum information measures in field theory at finite density and their implications. In this section we consider the entanglement entropy (EE) associated to the vacuum density matrix reduced to a region $V$ in space,
\begin{equation}
S(V)=-\Tr(\rho_V\log(\rho_V))\;\;,\;\; \rho_V=\Tr_{\bar{V}}(|0\rangle \langle 0|),
\end{equation}
where $\bar V$ is the complement of $V$, and $|0 \rangle$ is the vacuum state. 

Our original motivation in this direction came from~\cite{Swingle:2013zla}, who argued for a violation of the irreversibility of the RG in nonrelativistic models. In more detail, a Fermi surface in $d$ space-time dimensions leads to a logarithmic violation of the area law of EE,
\begin{equation}
S(V)\sim (k_F r)^{d-2}\log(k_F r)\,,
\label{eq:viol_area_law}
\end{equation}
where $V$ is a spherical region of radius $r$. At large $r$, this grows faster than the leading area law contribution,
\begin{equation}\label{eq:area}
S(V) \sim \frac{r^{d-2}}{\epsilon^{d-2}},
\end{equation}
that appears in local models, such as QFTs with UV fixed points. This argument towards a non-monotonic behavior is very suggestive, but various points need to be carefully understood.

First, the comparison is not well-defined in the continuum limit, because (\ref{eq:area}) is divergent. Making this precise requires a finite quantum information measure. Another issue is that the irreversibility of the RG in $d>2$ is not based on the area term (\ref{eq:area}); instead, the intrinsic quantities that decrease, $F$ and $A$~\cite{Casini:2012ei, Casini:2017vbe}, appear in subleading terms. Similar subleading terms have not been evaluated for finite density field theory in $d>2$. And finally, it is necessary to determine whether the chemical potential acts like a standard relevant operator whose effect becomes important at low energies or large radius as in (\ref{eq:viol_area_law}). This can be subtle, because a chemical potential modifies the structure of the ground state, which is in a different superselection sector than the zero charge vacuum. In particular, it could be clarifying to evaluate measures that can compare states in the same superselection sector.

Our goal is to analyze these points in the simplest setup of $d=1+1$ Dirac fermions at finite density which, as we shall see, is already nontrivial. Moreover, we would like to shed light on the competition between the mass and the chemical potential. The mass alone tends to give a trivial gapped state; but even with nonzero mass, the finite charge gives emergent massless fermions, as discussed in Sec.~\ref{subsec:dynamics}. This can produce long-range entanglement, and we want to characterize how this creation of entanglement occurs.

Recall that for a 2d conformal field theory of central charge $c$ and an interval of radius $r$, the leading term in the EE is~\cite{Calabrese:2009qy}
\be
S(r) = \frac{c}{3}\,\log\,\frac{r}{\epsilon}
\ee
with $\epsilon$ a short distance cutoff. The quantity
\begin{equation}\label{eq:cdef}
c(r)=r\frac{d S(r)}{dr}\,,
\end{equation}
is finite and is proportional to the intrinsic central charge $c$ at fixed points. It is also well-defined away from fixed points, in which case it decreases monotonically for unitary RG flows in relativistic theories~\cite{Casini:2004bw, Casini:2012ei}; this is the entropic version of the C-theorem. We will compute the finite quantity  (\ref{eq:cdef}) in the presence of a finite charge density, and use it to study potential violations of monotonicity. We also study RG flows in a fixed superselection sector, by comparing entropic $c$-functions with the same $k_F$ but different mass.

We perform numerical simulations using the real time approach~\cite{peschel2003,Casini:2009sr}; we work on the lattice of Sec.~\ref{subsec:latt_mod} and then take the continuum limit. Given that the theory is Gaussian, the eigenvalues of the reduced density matrix $\rho_V$ are determined by the eigenvalues of the correlation matrix $C_{ij}=\langle \psi_i^{\dag}\psi_j\rangle$ constrained to $V$ ($i,j\in V$). In terms of this matrix (see  (\ref{eq:Cij})), the entanglement entropy on the lattice reads
\begin{equation}\label{eq:Speschel}
S(V)=-\Tr[C\log(C)+(1-C)\log(1-C)]\,.
\end{equation}
We will now analyze two separate cases: chiral fermions and fermions at finite mass.

\subsection{Chiral fermions}\label{subsec:chiral_fermions}

For massless chiral Dirac fermions, we found a local unitary transformation (\ref{eq:localU}) that maps the theory with finite charge to the relativistic zero charge model. Both two-point functions $C_{ij}$ in the continuum limit have the same eigenvalues and, since the density matrix is Gaussian and completely determined by the two-point function, their respective density matrices also have the same eigenvalues. Therefore, quantum information measures that depend only on the eigenvalues of the reduced density matrix, such as the EE, are the same in both theories. This result is independent of the shape of the considered region $V$, so the statement also holds for an arbitrary number of intervals.

\begin{figure}[ht!]
    \centering
    \includegraphics[width=0.8\linewidth]{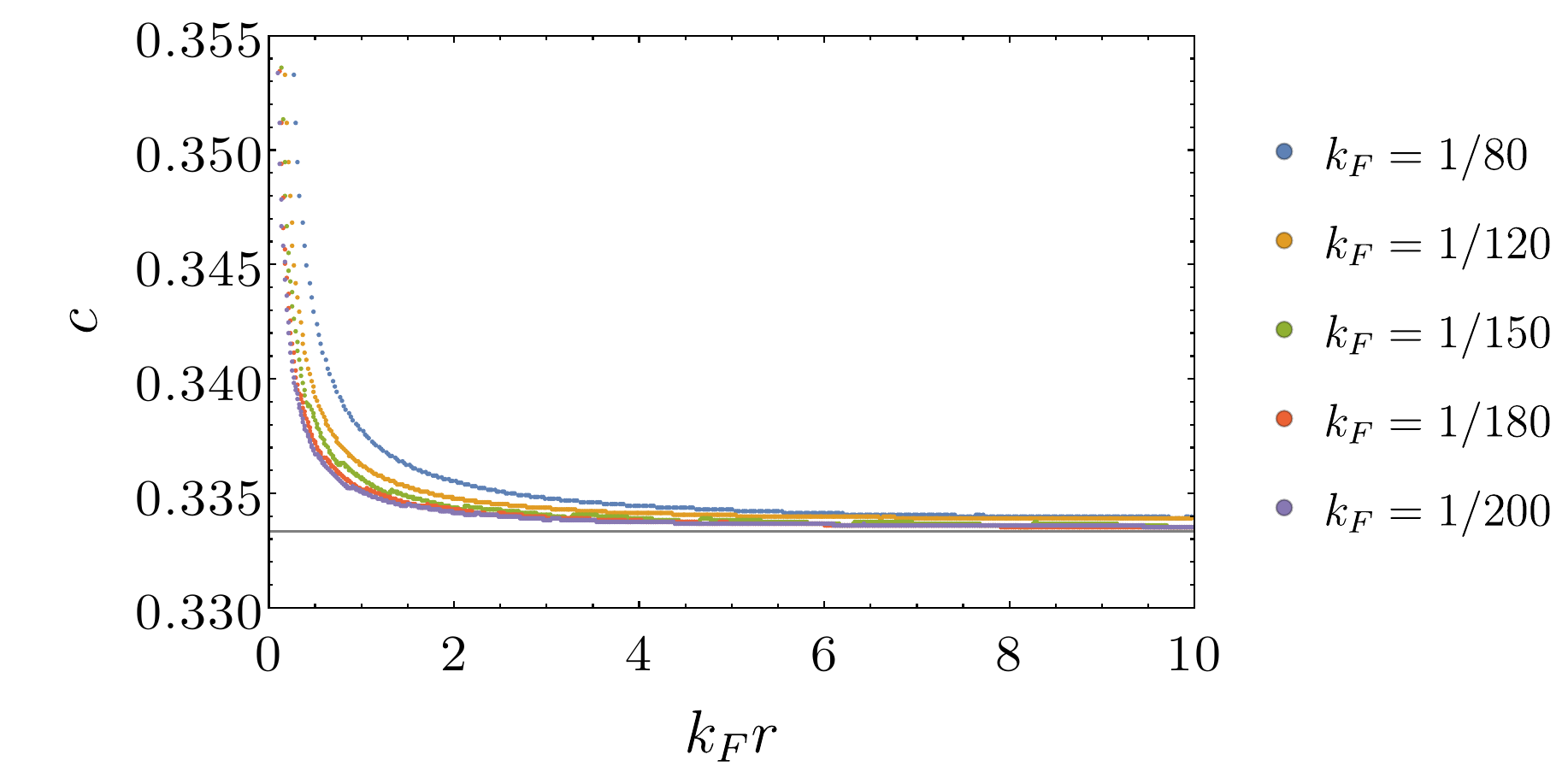}
    \caption{$c(k_Fr)$ function for a chiral fermion at finite density. For $k_F \to 0$ and $k_Fr$ fixed, the plots tend to $\frac{1}{3}$, as in a CFT. In this case, the correlation matrices were as large as $4000\times 4000$, since there are two spin degrees of freedom per site. Note the scale values in the $c$-axis.}
    \label{fig:EE_m0_simul}
 \end{figure}
 
\newpage
\noindent For an interval of length $r$, this implies
\be
c(r) = \frac{1}{3}
\ee
for chiral fermions at finite density. We provide a numerical check of this result in Fig.~\ref{fig:EE_m0_simul}, taking the continuum limit of the lattice model $k_F\to 0$, with $k_F r$ fixed. We present plots of $c(k_Fr)$ for several values of $k_F$.\footnote{To compute $c(k_Fr)=rS'(k_Fr)$, we used a mild discrete derivative that considers four consecutive points and reduces numerical fluctuations (see the numerical Appendix of \cite{Casini:2005rm}). More explicitly, $\partial_x \psi(x) \mapsto \psi_{n+\frac{1}{2}}=\frac{1}{4}(\psi_{n+1}+\psi_{n}-\psi_{n-1}-\psi_{n-2})$ for the $n$-th lattice site.} This is consistent with previous works~\cite{cardyconf,Ogawa:2011bz,Kim:2017xoh}, whose arguments were different from ours.  
 
Given this, it is interesting to explore measures that are not just sensitive to the eigenvalues of $\rho_V$, but also to the eigenvectors. This would distinguish chiral fermions with zero and finite density. We return to this point in Sec.~\ref{sec:relrenyi} below.

\subsection{Massive fermions at finite density}\label{subsec:massiveEE}

Let us now analyze the case of massive Dirac fermions. To develop analytical intuition, it is useful to consider first the asymptotic UV and IR limits, together with the ultra-relativistic and nonrelativistic behaviors. From the point of view of the EE and entropic $c$-function $c(r)$, the UV corresponds to $r \ll 1/m, 1/k_F$. The mass term is a standard relevant deformation, and hence its effect is negligible in the UV. On the other hand, as we discussed before, the charge density can be more subtle because it is changing the ground state. Assuming it also behaves like a relevant deformation, the UV limit should give $c(r) \to 1/3$. Our numerical results will show that this is indeed the case.

The functional form of $c(r)$ depends strongly on $m/k_F$. In the relativistic limit $m/k_F \to 0$, we expect a dependence close to that of chiral fermions in Sec.~\ref{subsec:chiral_fermions}. The nonrelativistic limit $m/k_F \gg 1$ is more nontrivial and interesting. From the point of view of the RG, the effects of the mass should set in first, at scales of order $mr \sim 1$, leading to a gapped state and hence to $c(r) \to 0$. However, even in this case, we expect at long distances $r \gg 1/m, 1/k_F$ to obtain nonzero entanglement from long range correlations of the light fermions in (\ref{eq:eff-nonrel}). Here we expect again $c(r) \to 1/3$. Therefore, the entropic $c$-function should exhibit a non-monotonic behavior in the nonrelativistic limit. 

\begin{figure}[ht!]
    \centering
    \includegraphics[width=0.9\linewidth]{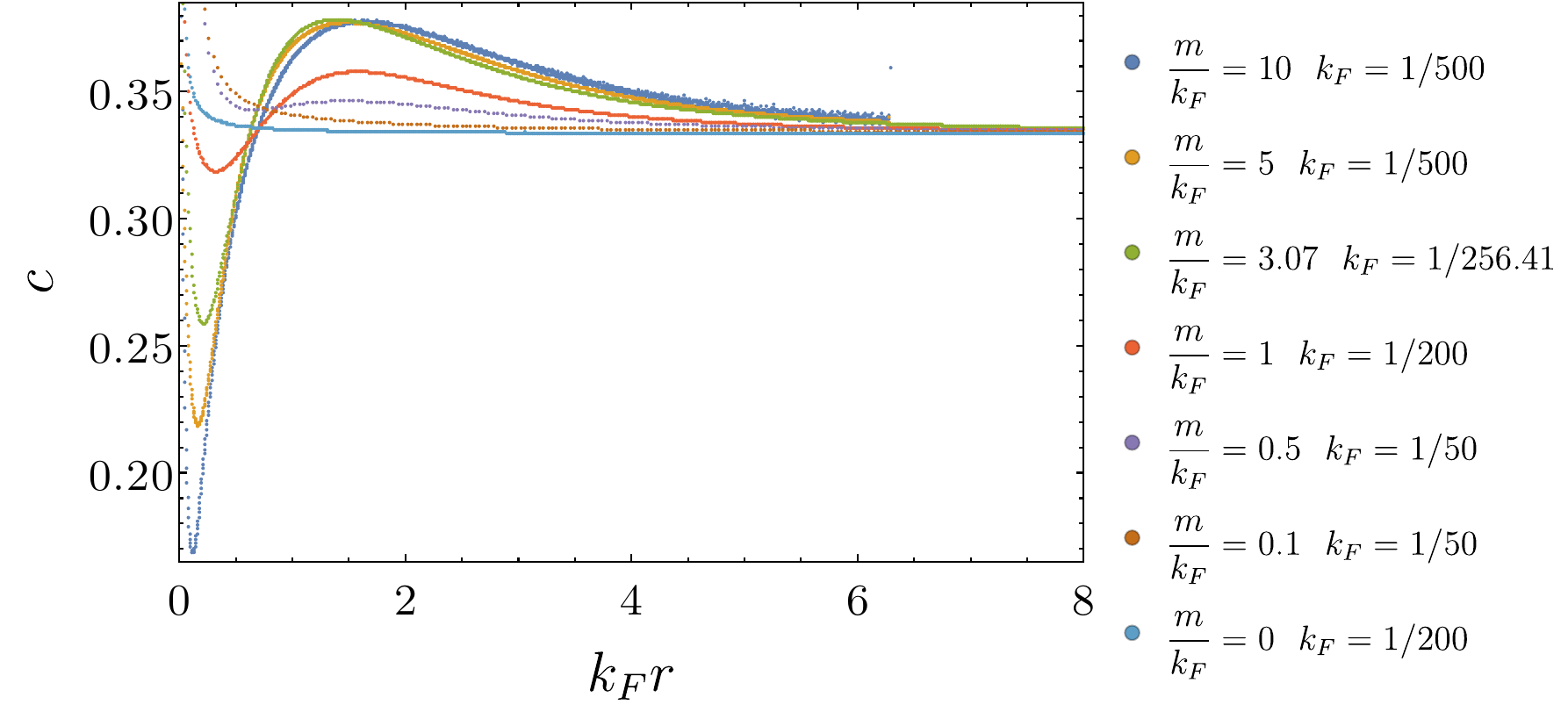}
    \caption{$c(k_Fr)$ function for different regimes $m/k_F$ for a Dirac fermion at finite density. In the limit $r \to 0$, the simulations tend to $\frac{1}{3}$, the expected value for the UV fixed point. In the IR limit, the simulations also tend to $\frac{1}{3}$, consistently with (\ref{eq:eff-nonrel}). Finally, the intermediate behavior is strongly affected by the ratio $m/k_F$.  The entropic $c$-function is a finite quantity that exhibits the non-monotonic behavior of the RG.}
    \label{fig:EE_c_all}
 \end{figure}

The numerical results are presented in Fig.~\ref{fig:EE_c_all}, which shows $c(k_Fr)$ for several values of $m/k_F$. We exhibit the ultrarelativisic $m/k_F \ll 1$ and nonrelativistic $m/k_F \gg 1$ limits, as well as intermediate cases with $m/k_F \sim 1$. 

The previous analytic considerations are verified by the numerical results. First, in all cases we find $c(r) \to 1/3$ in the UV limit $r \ll 1/m, 1/k_F$. This is the expected UV fixed point answer.
Secondly, in the IR limit ($mr\gg 1$ and $k_Fr\gg 1$) the plots also tend to $1/3$, the expected result from (\ref{eq:eff-nonrel}). This is the 2d version of the logarithmic violation of the area law (\ref{eq:viol_area_law}).
The intermediate behavior between these fixed points depends on $m/k_F$. In the relativistic regime $m\ll k_F$, the effects of the non trivial charge always dominate in the RG sense, and the behavior is similar to a chiral fermion.

In the non-relativistic regime, $c(r)$ rapidly decreases on scales $m r \sim 1$. This is consistent with the RG being dominated by the mass. For larger values of $r$, $c(r)$ increases, attaining a maximum at $k_Fr\sim1$, and finally asymptoting to $1/3$. The minimum and maximum in $c(k_Fr)$ reflect the competition between the operators $m \bar{\psi}\psi$ y $\mu_F \psi^{\dag}\psi$; the first tries to induce a mass gap and trivial entanglement, while the second (recall that $\mu_F > m$)  tries to induce long-range entanglement. The entropic $c$-function then provides a finite quantity that is measuring the creation of entanglement due to the finite density. 

Let us analyze this behavior in more detail. A nonrelativistic fermion at zero density is described by the Schrodinger action (\ref{eq:nonrel2}). It has a $U(1)$ symmetry $\t \psi \to e^{i \alpha} \t \psi$ with conserved particle number $N_e$; crucially, $N_e \ge 0$. This should be contrasted with the original $U(1)$ charge symmetry that has both positive and negative charges. The positivity of $N_e$ implies that the ground state is uniquely fixed as having zero charge at every (lattice) point. Therefore it factorizes, $|0\rangle = \prod_i |0 \rangle_i$ and the entanglement entropy on a finite region is trivial. This point was emphasized in~\cite{Hason:2017flq}; it can also be obtained directly as the $m \to \infty$ limit of the EE for a Dirac fermion. Now let us add some charge, so that the new ground state has $N_e >0$. There are many different ways to realize $N_e$, and the ground state is no longer a product state. With respect to $N_e$, there are excitations with charge excess or defect, particles and holes in condensed matter language. These excitations lead to nontrivial entanglement, similarly to what happens with particles and antiparticles in relativistic QFT. This creation of entanglement is detected by $c(r)$.

Finally, let us analyze the RG flow within the same charge sector. One way to characterize this is to compare the entropic $c$-functions for equal values of $k_F$ but different masses:
\be
\Delta c(r) =c(k_F r, m_1 r) -  c(k_F r, m_2 r) \,.
\ee
At long distance, this is changing the Fermi velocity $v_F$. Numerically we found that $\Delta c(r)$ also exhibits non-monotonic behavior.

To conclude, we have found non-monotonic behavior in the finite quantity $c(r)$, putting the considerations of~\cite{Swingle:2013zla}
 on a well-defined framework. Our results exhibit a breakdown of RG irreversibility in $1+1$ dimensions as measured by the entropic c-function, once Lorentz invariance is broken. As a cross-check, we have also verified that the strong subadditive inequality is always satisfied, as it should; in our case, this amounts to $S''(r)\geq 0$. In these flows, $c_{UV}=c_{IR}$, so the weak version of irreversibility, which states that $c_{UV} \ge c_{IR}$, is not violated. One could also ask whether other c-functions are monotonic. Since the UV and IR central charges are the same, this c-function would have to be a constant along the whole flow and hence insensitive to the two different relevant couplings $m$ and $\mu$. But this is quite implausible: at very short distances where we can focus on the effect of the relevant mass term, the consistent c-functions monotonically decrease with scale.\footnote{This includes the functional generalizations of the Zamolodchikov c-function in~\cite{Cappelli:1990yc}.} Given this, at larger scales they would then have to increase to asymptote to the same central charge in the IR. Therefore, generically we expect all c-functions to exhibit a non-monotonic behavior. Other information-theoretic quantities do exhibit monotonic behavior (most notably the relative entropy), and we discuss such measures in Sec.~\ref{sec:relrenyi}.

\section{Renyi entropies}\label{sec:REE}

In order to understand the entanglement spectrum in finite density QFT, 
in this section we analyze the Renyi entropies,\footnote{Besides giving the eigenvalues of $\rho_V$, the Renyi entropies are also important for the replica trick, where the limit $n \to 1$ gives the EE.}
\begin{equation}\label{eq:Sncn}
S_n(V)=\frac{1}{1-n}\log(\Tr(\rho_V^n))\;,\;c_n(r)=r\frac{dS_n(r)}{dr}\,.
\end{equation}
Ref.~\cite{Calabrese2010ParityEI} studied the $XY$ model on a 1d lattice and found a surprising behavior in the $S_n$, akin to Friedel oscillations in a metal. Their analytic prediction for the $S_n$ in the large distance limit $\log(2k_Fr)\gg n$ is
\begin{equation}
S_n(r)=\frac{n+1}{6n}\log\left(\frac{r}{\epsilon}\right)+A\,f_n \frac{\cos(2k_F r)}{(2k_Fr)^{\frac{2}{n}}}+\ldots,
\label{eq:osc_renyis}
\end{equation} 
with
\begin{equation}
f_n=\frac{2}{1-n}\left(\frac{\Gamma((1+n^{-1})/2)}{\Gamma((1-n^{-1})/2)}\right)^2\,.
\end{equation}
and $A=1$ in their case.

We would like to determine whether this phenomenon occurs more generally in finite density QFT in the continuum limit. In fact, it is possible to have oscillatory behavior on the lattice, but with the amplitude of the oscillations vanishing in the continuum; we will illustrate this in some of our numerical results.
The massive Dirac theory provides again a useful framework for understanding these points. Similarly to (\ref{eq:Speschel}), the Renyi entropies can be computed in terms of the two-point function $C$ restricted to a spatial region $V$, 
\begin{equation}\label{eq:SnC}
S_n(V)=\frac{1}{1-n}\Tr\left[\log(C^n+(1-C)^n)\right].
\end{equation}
The presence of charge density is crucial for such oscillatory behavior; however, this does not seem to be enough and the mass term should play a role as well. Indeed we argued in Sec.~\ref{subsec:chiral_fermions} that, in the massless limit, the zero density and finite density QFTs have the same spectrum for $\rho_V$. This predicts that the Renyi entropies should also be the same, and since the relativistic theory (a CFT) has no Friedel oscillations~\cite{Calabrese:2009qy, Casini:2005rm, Casini:2009sr}, they should not appear either at finite density.

With these motivations, in this section we evaluate explicitly the Renyi entropies and study the emergence of Friedel oscillations both in massless and massive theories.
\subsection{Numerical results}\label{subsec:REnumerics}
Let us begin with massless Dirac fermions on the lattice (\ref{eq:Hlattice}). This will provide an example of Friedel oscillations in lattice models that disappear in the continuum limit.
\begin{figure}[ht!]
    \centering
    \includegraphics[width=0.59\linewidth]{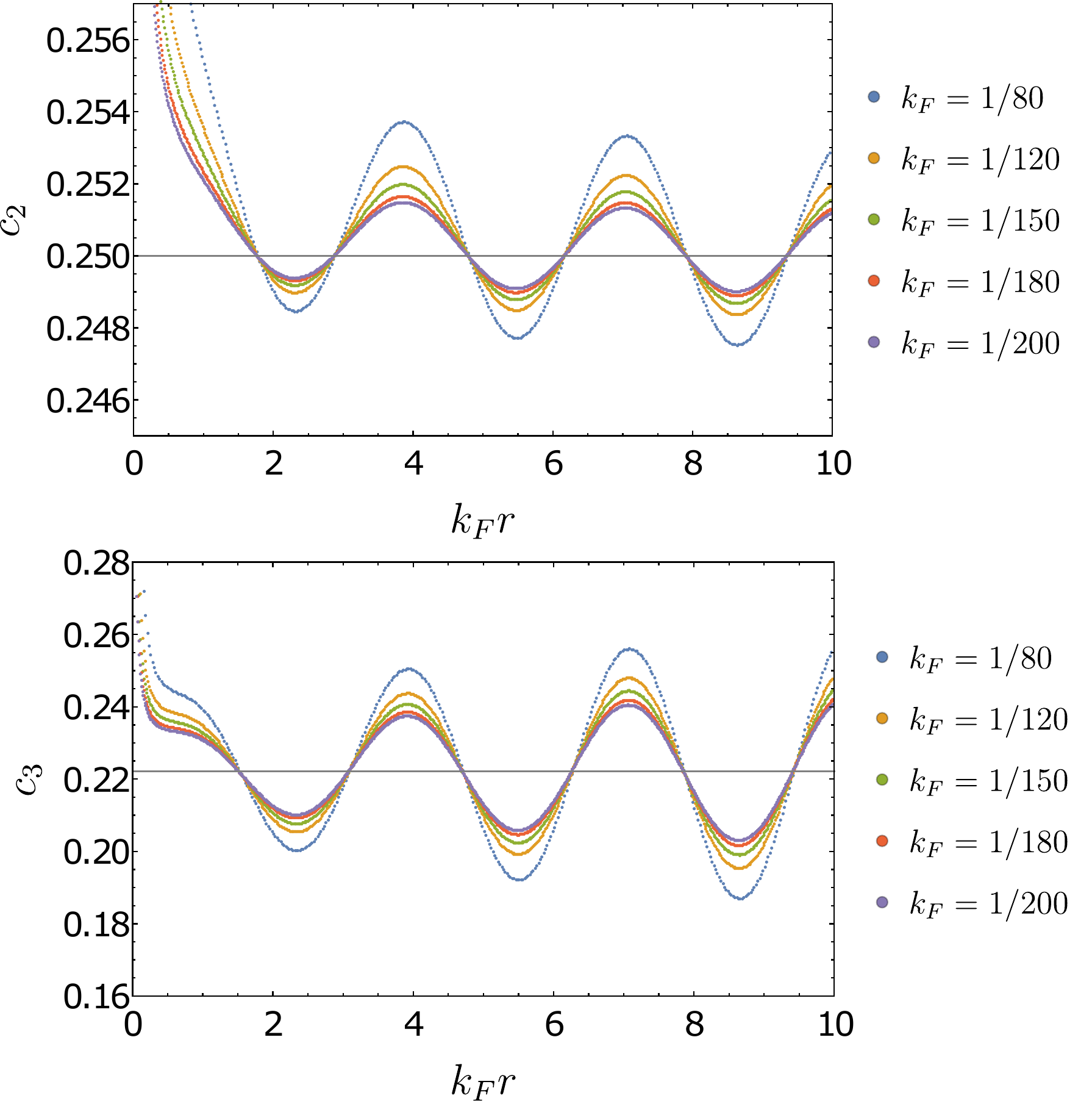}
    \caption{$c_2$ and $c_3$ functions for a chiral fermion at finite density. The continuum limit corresponds to $k_F \to 0$ for $k_Fr$ finite. The amplitudes of oscillations vanish in this limit, and the results converge to (\ref{eq:relcn}). Note the scale values in the $c_n$-axis.}
    \label{fig:Cn_masless}
 \end{figure}
\newpage

Numerical results for Renyi entropies are presented in Fig.~\ref{fig:Cn_masless}. They feature Friedel oscillations; however, we find that the amplitude of such oscillations goes to zero in the continuum limit, and illustrate this in the figure by plotting curves of decreasing $k_F$ in units of the lattice spacing.  This vanishing of the oscillation amplitude is
in agreement with the argument in Sec.~\ref{subsec:chiral_fermions} that the result should be the same as in the relativistic fixed point
\be\label{eq:relcn}
c_n= \frac{n+1}{6n}\,,
\ee
with $c_n$ defined in (\ref{eq:Sncn}). Reintroducing the lattice spacing $a$ that was fixed before to $a=1$, the amplitude $A$ introduced in (\ref{eq:osc_renyis}) vanishes as $(k_F a)^2$ when the lattice spacing $a \to 0$.
We show the convergence to the continuum limit $k_F \to 0$ keeping $k_Fr$ fixed, as well as the prediction (\ref{eq:relcn}).

\begin{figure}[ht!]
    \centering
  \includegraphics[width=0.75\linewidth]{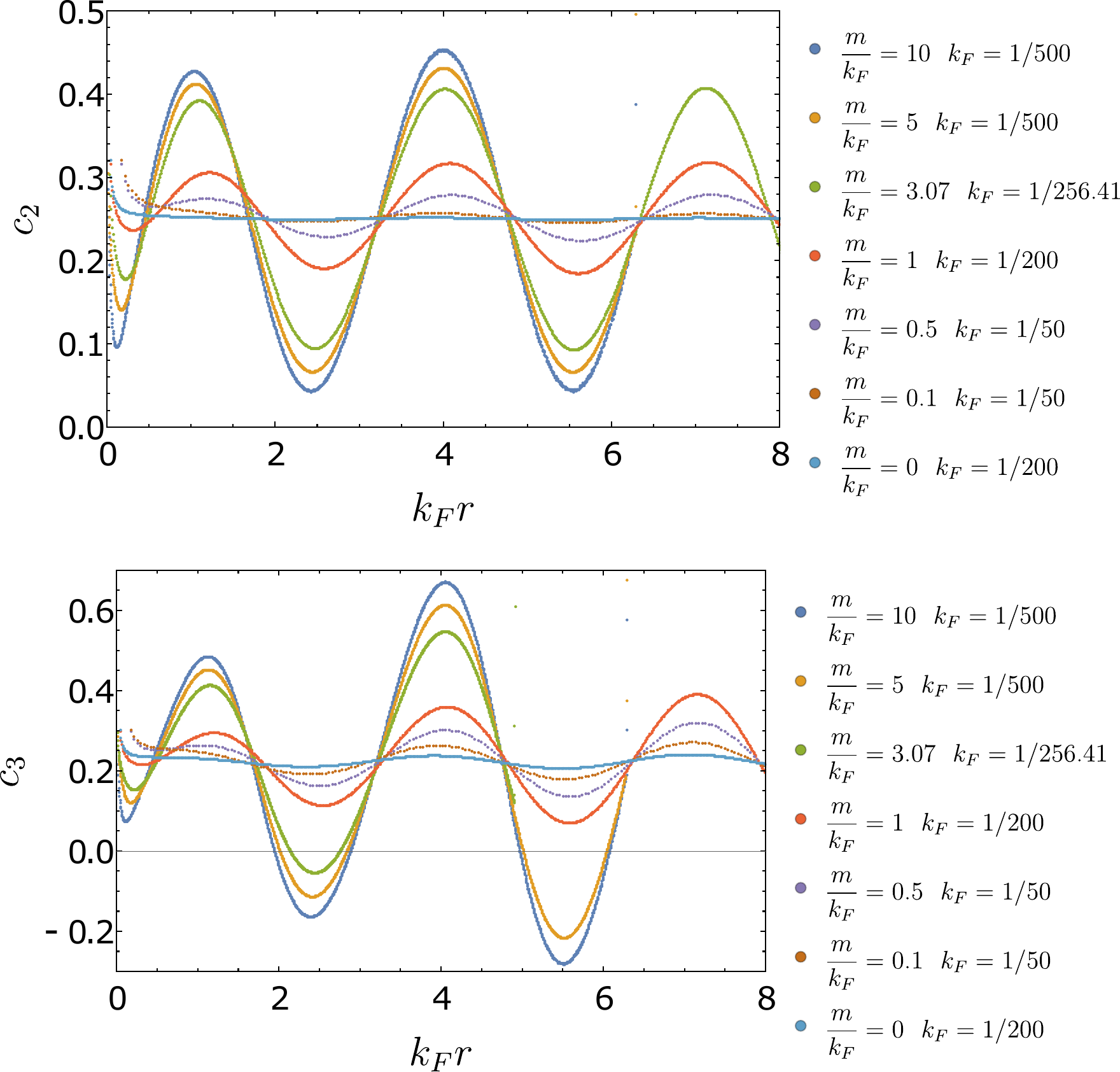}
    \caption{$c_2$ and $c_3$ functions for all regimes of Dirac fermions at finite density. In all of them, except for $m=0$, there are Friedel oscillations of period $\pi$ as a function of $k_Fr$.}
    \label{fig:Sn_all}
 \end{figure}

Next, we consider the massive case. Fig.~\ref{fig:Sn_all} shows our numerical results for $c_2$ and $c_3$ in different regimes of $m/k_F$.

\begin{figure}[ht!]
    \centering
    \includegraphics[width=0.95\linewidth]{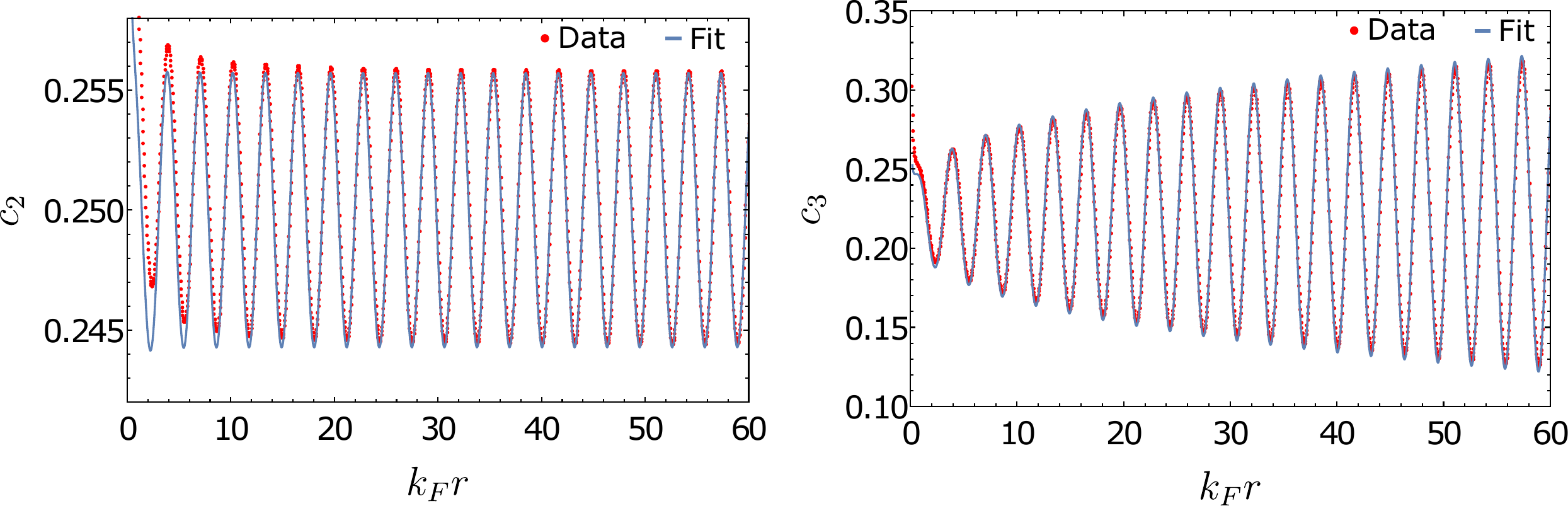}
    \caption{Fits of $c_2$ and $c_3$ for $m/k_F=0.1$ with $k_F=1/50$. The fits were used following the expression (\ref{eq:osc_renyis}) and $c_n(r)=rS'_n(r)$. The free parameter of every fit is the amplitude $A$.}
    \label{fig:ajuste_smallm}
 \end{figure}

For $n>1$ we find Friedel oscillations in the continuum limit, not only when $m \gg k_F$ but in other ranges of $m/k_F$ as well. The oscillations have a constant mean value (\ref{eq:relcn}) due to the long range contributions coming form the light fermions of the low energy theory (\ref{eq:eff-nonrel}). Moreover, the long-distance dependence (\ref{eq:osc_renyis}) also fits well away from the nonrelativistic limit. This is presented in Fig.~\ref{fig:ajuste_smallm} for $c_2$ and $c_3$ in the $m/k_F \ll 1$ limit; the agreement with (\ref{eq:osc_renyis}) is excellent.

\subsection{Friedel oscillations and defect OPE}\label{subsec:friedel}

Friedel oscillations have been observed before in lattice systems with Fermi surfaces, such as~\cite{Calabrese2010ParityEI}; this reference also provided an analytic calculation that leads to (\ref{eq:osc_renyis}), based on the behavior of  discrete Toeplitz matrices at large separation. Here we have found these effects in continuum QFT, and we would like to understand their physical origin. In~\cite{Swingle_2013} it was suggested that they arise from localized terms on the defect that defines the Renyi entropy, by developing an argument given originally in~\cite{Cardy2010UnusualCT}. In this subsection we will show how such phenomenon can be understood using the defect operator product expansion (OPE). 

The Renyi entropy $\tr \rho_V^n$ for a region $V$ can be thought of as the partition function in the presence of a codimension 2 defect operator $\Sigma_V^n$, called a twist operator, that implements the joining of the different replicas,
\be\label{eq:defSigma}
\tr \rho_V^n=\langle \Sigma_V^n\rangle\,.
\ee
In $d=1+1$ space-time dimensions, the defect is just two points (the endpoints of the intervals); these local operators are the twist and anti-twist introduced first in~\cite{Cardy:2007mb} (see~\cite{Calabrese:2009qy} for a review). In higher dimensions, $\Sigma_V^n$ is an extended codimension 2 defect. 

We are interested in studying the behavior of a local operator $\mathcal O$ near the defect, and we assume for now we have a CFT. We will discuss  the application to our case momentarily. Let us choose a local coordinate system where $\vec y$ are $d-2$ coordinates tangential to the defect, while $x^a$ are the two coordinates normal to the defect (located locally at $x^a=0$). In the limit where the operator  $\mathcal O$ is very close to the defect, we can use a defect OPE to write it in terms of operators $\hat {\mathcal O}$ localized at the defect:
\be\label{eq:defectOPE1}
\mathcal O(x^a, \vec y) \sim \sum_k\,b_k |x^a|^{\hat \Delta_k - \Delta} \hat{\mathcal O}_k( \vec y)\,,
\ee
as discussed in~\cite{CastroAlvaredo:2011du, Levi:2012zj, Bianchi:2015liz, Billo:2016cpy, Gadde:2016fbj}. Intuitively, the OPE coefficient $b_k$ measures the strength with which $\mathcal O$ induces the operator $ \hat{\mathcal O}_k$ localized on the defect. Similarly, the defect can be expanded in terms of defect-localized operators, and this expansion is expected to exponentiate,
\be\label{eq:exp}
\Sigma_V^n \sim e^{- \int d^{d-2}y \,b_k^\Sigma\,\epsilon^{ \hat \Delta_k-(d-2)}\, \hat{\mathcal O}_k( \vec y)}\,,
\ee
by arguments similar to those in~\cite{Bousso:2014uxa}.\footnote{For $d=2$, the twist operator is point-like so we expect an expansion in terms of local operators and not with an exponential. This is not important for our argument below, which only uses the leading contribution in this expansion.}
For dimensional reasons, we have included the factor $\epsilon^{ \hat \Delta_k-(d-2)}$, where $\epsilon$ is a short distance cutoff that defines a tubular region around $\Sigma_V^n$. The OPE coefficients are fixed by independent calculations of correlation functions in the conical geometry, such as
\be
\langle \Sigma_V^n \mathcal O(x^a, \vec y) \rangle\;,\;\langle \Sigma_V^n \mathcal O(x^a, \vec y) \mathcal O(x'^a, \vec y') \rangle\;,\;\ldots
\ee
see for instance~\cite{Guimaraes:1994sw} for a scalar field example.

We won't need the explicit OPE coefficients, but note that in general we expect $b_k^\Sigma \sim n-1$ when $n \to 1$ since
\be
\langle \Sigma_V^n \mathcal O(x^a, \vec y) \rangle \sim n-1
\ee
(in this limit the defect is becoming trivial). Typically all operators allowed by symmetries to have a nonzero expectation value in the conical geometry will contribute to the OPE, and the dynamics is dominated by the one with smallest scaling dimension. Furthermore, plugging (\ref{eq:exp}) into (\ref{eq:defSigma}) and writing explicitly the euclidean path integral, we have
\be\label{eq:actiontrrho}
\tr \rho_V^n \sim \int D\phi\,e^{-S[\phi]- \int d^{d-2}y \,b_k^\Sigma\,\epsilon^{\hat \Delta_k-(d-2)}\, \hat{\mathcal O}_k( \vec y)}\,.
\ee
Therefore the $\hat{\mathcal O}_k$ that enter the defect OPE appear as contributions to the action localized on the defect.

In order to understand more explicitly the kinds of defect operators $\hat{\mathcal O}_k$ that can appear, let us specialize to $d=2$ and a single interval. We work with complex coordinates $w$; denote the endpoints in the complex $w$ plane by $(u,v)$. The replicated manifold $\mathcal M_n$ corresponds to an $n$-cover of the complex plane, branched over $u$ and $v$. It can be mapped to $\mathbb C$ via the uniformization map~\cite{Calabrese:2009qy}
\be\label{eq:transf1}
z= \left(\frac{w-u}{w-v} \right)^{1/n}\,,
\ee
where $w \in \mathcal M_n$ and $z \in \mathbb C$. Under such a conformal transformation, a scalar primary operator of dimension $\Delta$ transforms as
\be
\mathcal O(w) = \left(\frac{dz}{dw} \right)^\Delta\,\mathcal O'(z)\,.
\ee
Taking the OPE limit $w \to u$ and 
expanding the conformal factor, we find
\be\label{eq:OOhat}
\mathcal O(w) \sim |w-u|^{\frac{\Delta}{n}- \Delta}\, \mathcal O'(0) + \ldots
\ee

Given the OPE expansion (\ref{eq:defectOPE1}), we interpret the right hand side as a defect operator $\mathcal O'(0)$ of fractional dimension $\Delta/n$ induced by the bulk operator $\mathcal O(w)$. Finally, recalling (\ref{eq:exp}) and (\ref{eq:actiontrrho}), we conclude that a conformal primary of dimension $\Delta$ in the bulk, with nonzero expectation value on the conical geometry, will induce a localized term containing an operator of dimension $\Delta/n$. This fact was suggested before in~\cite{Cardy2010UnusualCT} based on related behavior of boundary states and lattice effects near conical singularities. Here we have understood it from the point of view of the defect OPE.

Let us apply these results to our setup. We work at long distance $k_F r \gg 1$. The low energy theory (\ref{eq:eff-nonrel}) is a CFT with two chiral fermions $\t \psi_L$ and $\t \psi_R$. The ultraviolet cutoff for this description is $k_F$, so the natural value for the small distance parameter above is $\epsilon \sim 1/k_F$. The lowest dimension operators with nonzero expectation value $\langle \Sigma_V^n \mathcal O(x^a, \vec y) \rangle$ are constructed from fermion bilinears. Besides $\t \psi_L^\dag \t \psi_L$ and $\t \psi_R^\dag \t \psi_R$, we have the $2 k_F$ density operator
\be
\rho_{2 k_F} = e^{2ik_Fx^1}\t \psi^{\dag}_L \t\psi_R+\text{h.c.}
\ee
that appears in the conserved current $\bar \psi \gamma^\mu \psi$ of the microscopic theory. The two-point function for $\rho_{2 k_F}$ exhibits the familiar Friedel oscillations in Fermi liquids. When this operator is very close to an endpoint of our entangling interval, we can still apply the transformation (\ref{eq:transf1}) that leads to (\ref{eq:OOhat}); $e^{2ik_Fx^1}$ appears as a slowly-varying overall prefactor. Therefore, the $2k_F$ operator of $\Delta=1$ leads to localized operators $\hat \rho_{2k_F}$ at the endpoints of the interval, of dimension $\hat \Delta=1/n$ and strength $n-1$ for $n \to 1$. The leading contribution in this limit is
\bea
\tr \rho_V^n &\sim& \int D\t \psi \,e^{-S_{CFT}[\t \psi]} \left( (n-1) k_F^{-1/n}\, (\hat \rho_{2k_F}(u)\right)\left( (n-1) k_F^{-1/n}\, (\hat \rho_{2k_F}(v)\right) \nonumber\\
& \sim& (n-1)^2 \frac{\cos(2k_F r)}{(k_F r)^{2/n}}\,.
\eea
This identifies the Renyi entropy Friedel oscillations as arising from fractional $2k_F$ operators localized at the endpoints of the entangling region.\footnote{This was proposed before by~\cite{Swingle_2013} based on the observation of localized operators given in~\cite{Cardy2010UnusualCT}.} Our approach also explains why the oscillations vanish when $n=1$, a point that was not clear in previous works. 

\begin{figure}[ht!]
    \centering
    \includegraphics[width=0.65\linewidth]{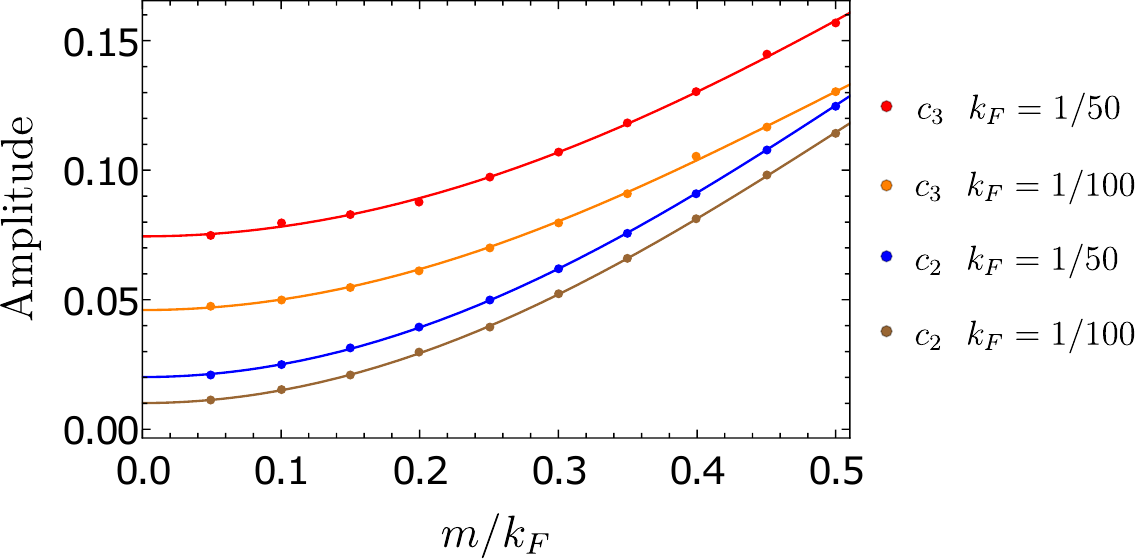}
    \caption{Fits $y=a+b\left(\frac{m}{k_F}\right)^2+c\left(\frac{m}{k_F}\right)^4$ for the amplitude $A$  in (\ref{eq:osc_renyis}), for $c_2$ and $c_3$, with $k_F=1/50$ and $k_F=1/100$. For a given $n$, one curve is under another one when diminishing $k_F$; additional values of $k_F$ should be included in order to take the continuum limit. But already at this order, we find that $b$ is the same for the two values of $k_F$, suggesting we are not far from the continuum limit. We obtain $b=0.49$ for $n=2$ and $b=0.38$ for $n=3$.}
    \label{fig:ajuste_smallm2}
 \end{figure}

In the relativistic limit $m \to 0$ the Friedel oscillations in the Renyi entropy vanish because the eigenvalues of the density matrix are the same as in the zero charge theory in the continuum.\footnote{Furthermore, these oscillations are not expected in the Renyi entropies in the relativistic case, given the reflection positivity inequalities of~\cite{Casini:2010nn}. We thank H.~ Casini for pointing this out.} The amplitude $A$ of the oscillations should then have a perturbative expansion in powers of $(m/k_F)^2$. We have checked this numerically, as seen in Fig.~\ref{fig:ajuste_smallm2}. This suggests that it might be possible to perform the perturbative expansion analytically, for instance using bosonization techniques~\cite{Casini:2005rm, Aristov_2002}.

Finally, we note that the defect OPE approach is also general, and could be applied to interacting theories at finite density, such as those that arise in non-Fermi liquids. Besides the conserved current, other fermion bilinears with nontrivial anomalous dimension could contribute Friedel oscillations to the Renyi entropy, and this could be an interesting measure for their nontrivial scaling.

\section{Mutual information}\label{sec:mutual}

The previous sections dealt with information measures associated to a single connected region (an interval). On the other hand, measures based on two non-intersecting regions $A$ and $B$ are very interesting to study because they can detect correlations between $A$ and $B$. In this section we will analyze the mutual information, defined in terms of EE entropies as
\be\label{eq:Idef}
I(A,B)= S(A) + S(B) - S(A \cup B)\,;
\ee
we also consider the Renyi version, given by
\be\label{eq:In}
I_n(A,B)= S_n(A) + S_n(B) - S_n(A \cup B)\,.
\ee

There are various properties of the mutual information that motivate its study.  First, it measures the total amount of correlations between $A$ and $B$. Given observables $M_A$ and $M_B$ with connected correlation function $C(M_A, M_B) = \langle M_A \otimes M_B \rangle - \langle M_A \rangle \langle M_B \rangle$, we have the bound~\cite{Wolf:2007tdq}
\be\label{eq:Ibound}
I(A,B) \ge \frac{C(M_A, M_B)}{2 ||M_A||^2 ||M_B||^2}\,.
\ee
The mutual information is then a good measure for the total correlations, some of which can be missed by looking at specific correlation functions that could be small. From the point of view of QFT, another important property is that divergent boundary contributions cancel out so, unlike the EE, (\ref{eq:Idef}) is well-defined in the continuum limit.

For the purpose of this work, a key property is that $I_n(A,B)$ admits an operator product expansion (OPE) in the limit where the sizes $r_A, r_B$ of $A$ and $B$ are much smaller than the distance $L$ between them~\cite{Cardy.esferaslejanas}
\be\label{eq:limitI}
\frac{r_A}{L} \ll 1\,,\frac{r_B}{L} \ll 1\,.
\ee
As we review below in Sec.~\ref{subsec:OPEI}, the basic idea is that the twist operator $\Sigma_A^n$ that implements $\tr(\rho_A^n)$, can be expanded into a basis of local operators when we are far away from $A$; and similarly for $\Sigma_B^n$. The connected part of the two-point function $\langle \Sigma_A^n \Sigma_B^n \rangle$, which determines $I_n$, can then be expanded into a sum of correlation functions for operators between $A$ and $B$. For a CFT when the leading contribution comes from exchange of an operator of dimension $\Delta$, this gives
\be
I \sim \left(\frac{r_A r_B}{L^2} \right)^{\Delta}\,.
\ee

We analyze how this is modified at finite density, finding that the mutual information extracts detailed information about the dynamics. One of our main results is that a Fermi surface and its long range correlations modifies the mutual information already at leading order by introducing oscillating terms. The physical origin of these oscillations turns out to be different from the Friedel oscillations above, and this provides a new quantum information probe for (non)-Fermi liquids.

\subsection{Analysis via the OPE expansion of the mutual information}\label{subsec:OPEI}

For the regions $A$ and $B$ we take two intervals of length $r$, separated by a distance $L$, and focus on the limit (\ref{eq:limitI}). We have argued above that the spectrum of the density matrix in the massless case is the same as that with zero charge, so we focus on $m \neq 0$ with a finite charge density.
We first evaluate the mutual information analytically using the OPE expansion of~\cite{Cardy.esferaslejanas}, and in the next subsection we compare with numerical computations.

We recall from the discussion in Sec.~\ref{subsec:friedel} that Renyi entropies are implemented in terms of twist operators $\Sigma$. The Renyi entropy of the union $A \cup B$ is then proportional to the two-point function 
\be\label{eq:SigmaSigma}
\tr \rho_{A \cup B}^n=\langle \Sigma_A^n \Sigma_B^n\rangle\,.
\ee
The idea of~\cite{Cardy.esferaslejanas} is that far away from the region $A$ the twist can be expanded as
\be\label{eq:Sigmaope}
\Sigma^n_A=\sum_{\{k_j\}} C^A_{\{k_j\}}\, \prod_{j=0}^{n-1} \mathcal O_{k_j}(r_A^j)\,,
\ee
where $\mathcal O_{k_j}$ is an operator in the $j$-th copy, and $k_j$ is an index that determines the type of operator. The OPE coefficients $C^A_{\{k_j\}}$ are obtained from expectation values of operators far away from $A$ in the presence of the defect,
\be\label{eq:CA}
\lim_{r \to \infty}\, \langle \Sigma_A^n\,\prod_{i=0}^{n-1} \mathcal O_{k'_i}(r) \rangle =\lim_{r \to \infty}\, \sum_{\{k_j\}} C^A_{\{k_j\}}\,\prod_{j=0}^{n-1} \langle  \mathcal O_{k'_j}(r)  \mathcal O_{k_j}(r_A^j) \rangle\,.
\ee
The leading contribution always comes from the identity operator, and can be normalized to $\tr \rho_{A \cup B}^n\Big|_{\mathcal O= \mathbf 1}=1$. If the next dominant contribution comes from an operator $\mathcal O_{k_j}$, expanding $\log (\tr \rho_{A \cup B}^n)$ gives a Renyi mutual information
\be\label{eq:InOPE2}
I_n(A,B) \approx \frac{1}{1-n}\,\sum_{\{k_j\}}\, \sum_{j=0}^{n-1}\,C^A_{\{k_j\}}\,C^B_{\{k_j\}}\, \langle \mathcal O_{k_j}(r_A^j) \mathcal O_{k_j}(r_B^j) \rangle\,.
\ee

In our case, the lowest dimension operators that contribute to (\ref{eq:Sigmaope}) are fermion bilinears. The independent bilinears in even $d$ space-time dimensions can be chosen as
\be\label{eq:bilinears}
\bar \psi_\alpha \psi_\beta\;:\;\;\bar \psi \psi\;,\; \bar \psi \gamma^\mu \psi\;,\;\bar \psi \psi\;,\; \bar \psi [\gamma^\mu, \gamma^\nu] \psi\;,\;\ldots\;,\;\bar \psi \gamma^{*d} \psi\,,
\ee
with $\gamma^{*d}= \gamma^0 \ldots \gamma^{d-1}$
There are two kinds of contributions to (\ref{eq:Sigmaope}): the two fermions on different replicas, or both on the same replica. So at leading order,
\be\label{eq:SigmaA-Dirac}
\Sigma^n_A \sim  \prod_{j=0}^{n-1} \prod_{j'\neq j}\,C^{A\,\alpha \beta}_{jj'}\, \bar \psi_\alpha(r_A^j)  \psi_\beta(r_A^{j'})+ \prod_{j=0}^{n-1}\,C^{A\,\alpha \beta}_{jj}\, \bar \psi_\alpha(r_A^j)  \psi_\beta(r_A^j)\,.
\ee
The $C_{jj}^A$ coefficients are proportional to $n-1$ for $n \to 1$; they contribute to the Renyi mutual information but, since they appear squared in (\ref{eq:InOPE2}), they do not contribute to the limit $n \to 1$ that gives the mutual information~\cite{Agon:2015ftl, Agon:2015twa, Chen:2017hbk}. In contrast, the coefficients $C_{jj'}^A$ involving different replicas give a nonzero contribution to the mutual information. For Dirac fermions in $d$ dimensions with $m=k_F=0$, the scaling dimension is $\Delta=(d-1)/2$. From (\ref{eq:CA}), the OPE coefficients scale like $C^A \sim r^{d-1}$; plugging (\ref{eq:SigmaA-Dirac}) into (\ref{eq:SigmaSigma}), gives a scaling
\be
I_n \sim \left(\frac{r_A r_B}{L^2} \right)^{d-1}\,.
\ee
Recall that the regions have characteristic sizes $r_A$ and $r_B$, and $L$ is the distance between them.
Furthermore, it turns out that only the bilinear $\bar \psi \gamma^\mu \psi$ in (\ref{eq:bilinears}) contributes to the operator exchange between the two regions~\cite{Chen:2017hbk}.

In the massive case with zero charge, correlation functions are exponentially suppressed at long distances, and the same exponential suppression is seen in the mutual information. Once $k_F \neq 0$, however, the Fermi surface gives rise to long-range correlations as we found above. In $d$ spacetime dimensions, a Fermi surface behaves like a collection of massless $1+1$-dimensional fermions, one for each patch of the Fermi surface. Therefore, we expect to see a scaling behavior associated to a 2d CFT. This dynamical change in the effective dimensionality is called ``hyperscaling violation''~\cite{Fisher:1986zz, Huijse:2011ef, Dong:2012se}. Here we consider the case $d=1+1$, for which there is no hyperscaling violation; previous work in higher dimensions includes~\cite{swingle2012renyi}.

The leading effect at finite density is associated to oscillating terms in the fermionic correlators. At large distances, these can be obtained using the low energy theory (\ref{eq:eff-nonrel}) in terms of the left and right movers $\t \psi_L,\, \t \psi_R$; they have a linear dispersion relation with velocity $v_F$. The bilinears (\ref{eq:bilinears}) exhibit oscillating terms when expressed in terms of these low energy fields. For instance, the charge density contains a term
\be\label{eq:exchange}
 \bar \psi(x) \gamma^0 \psi(x) \,\supset\, e^{2i k_F x}\, \t \psi^\dag_L \t \psi_R + h.c.
\ee
These contributions have the same scaling behavior as the non-oscillating terms that do not mix $\t \psi_L$ and $\t \psi_R$. Therefore, the (Renyi) mutual information in the long distance OPE limit is
\be\label{eq:Inprediction}
I_n \sim \frac{r^2}{L^2} \left( a_n + b_n \,\cos(2 k_F L+ \phi_n) + \ldots\right)\,,
\ee
with $a_n, b_n$ some $O(1)$ constants.

The main conclusion from this OPE calculation is that the mutual information detects the presence of a Fermi surface already at leading order in $r^2/L^2$ via oscillating terms. The charge density or Fermi momentum can be read off from the frequency of oscillations. This applies both to the mutual information and to the Renyi mutual information. Note that the origin of these oscillations is distinct from the Friedel oscillations in the Renyi entropies obtained in Sec.~\ref{sec:REE}. Unlike the Renyi entropies, here we find oscillations in the mutual information already at $n=1$; they arise from long-distance correlations of fermion bilinears such as (\ref{eq:exchange}), to which the mutual information is sensitive at leading order. In the Renyi entropy case, the oscillations were coming from boundary operators localized at the endpoints; these boundary effects cancel in the mutual information.

This is one of our main results, and we verify it numerically in the next subsection. We believe that these properties make the mutual information and interesting probe for finite density dynamics. While the present work focuses on free theories, we can also use the OPE for interacting Fermi surfaces with non-Fermi liquid behavior. Once the fermionic quasiparticles acquire an anomalous dimension, their scaling dimension $\Delta_f$ becomes greater than $1/2$. Based on the previous OPE analysis, we expect a mutual information of the form
\be
I \sim \left(\frac{r^2}{L^2}\right)^{2\Delta_f} \left( a_1 + b_1 \,\cos(2 k_F L+ \phi_1) + \ldots\right)\,.
\ee
Therefore the mutual information would detect both the non-Fermi liquid scaling dimension as well as the Fermi momentum. The behavior of the Renyi mutual information is different, because the charge density has protected dimension $\Delta_J=1$ and contributes via the $C_{jj}^A$ in (\ref{eq:SigmaA-Dirac}). This term vanishes when $n \to 1$ as we discussed before. Therefore we expect to see two different scaling dependences in the Renyi mutual information. More generally, it would be interesting to consider experimental applications of these results.

\subsection{Numerical results}\label{subsec:Inumerics}

We now compute the mutual information numerically; the approach is similar to that of Secs.~\ref{sec:EE} and~\ref{sec:REE}. The new ingredient is the correlator matrix $C_{ij}$ with indices restricted to the disconnected region $A \cup B$. The mutual information 
(\ref{eq:Idef}) is then evaluated in terms of (\ref{eq:Speschel}); similarly, for $I_n$ we use (\ref{eq:SnC}). We fix equal lengths $r$ for the two intervals, and vary their distance $L$, focusing on $L \gg r$.

\begin{figure}[htb!]
    \centering
    \includegraphics[width=0.6\linewidth]{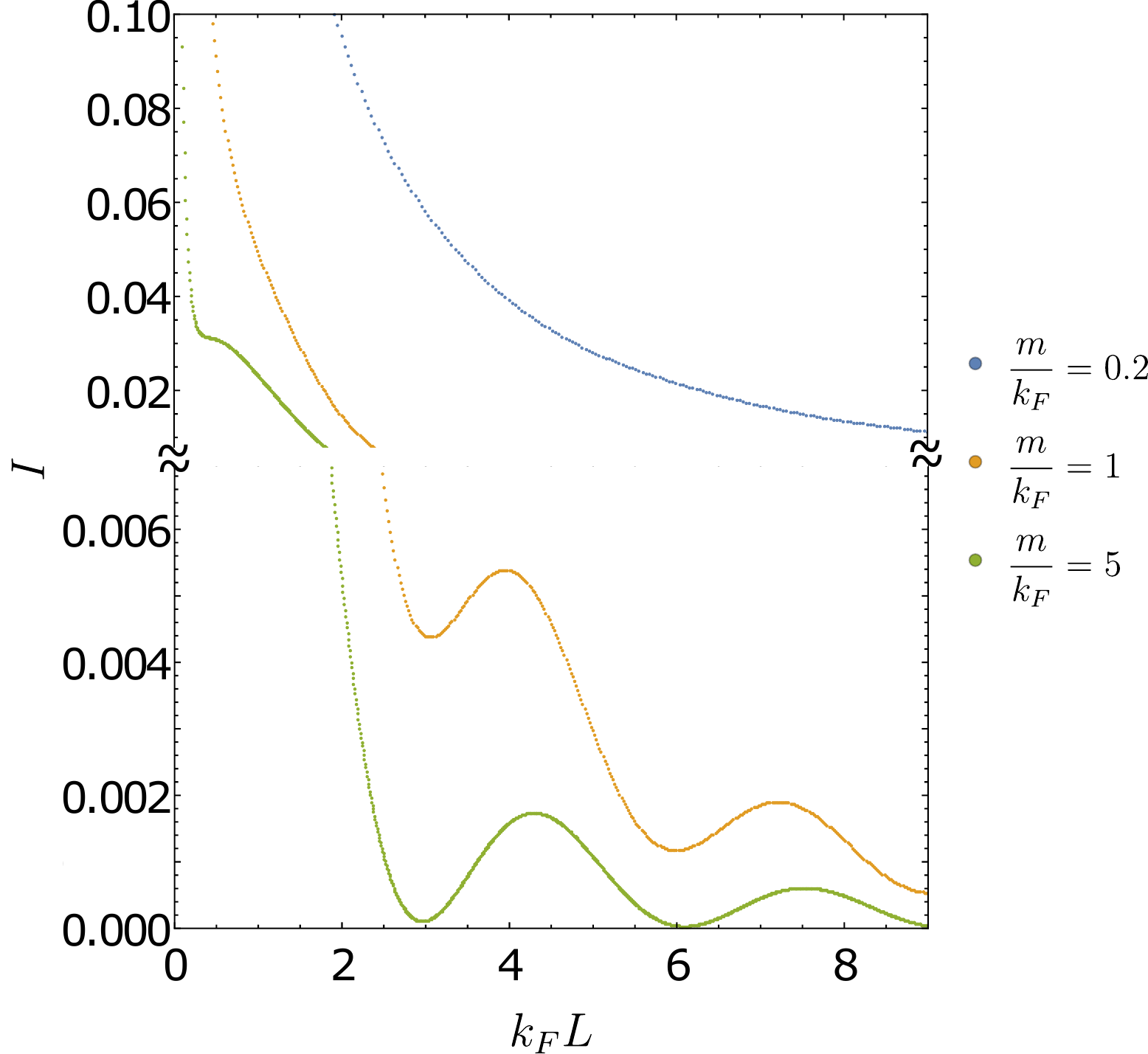}
    \caption{$I(k_FL)$ for fixed $r$ for massive Dirac fermions at finite density. The lengths $r$ were $k_Fr=2$, $k_Fr=0.5$ and $k_Fr=0.2$ for $\frac{m}{k_F}=0.2$, $\frac{m}{k_F}=1$ and $\frac{m}{k_F}=5$ respectively. Oscillations of period $\pi$ as function of $k_FL$ are present. Their amplitude vanishes when $m/k_F \to 0$.}
    \label{fig:I_all}
 \end{figure}
   
Let us begin with the mutual information. The numerical results for $I(k_FL)$ at fixed $r$ are shown in Fig.~\ref{fig:I_all}. We compute three curves, corresponding to the ultra-relativistic limit $m/k_F \ll 1$, the nonrelativistic limit $m/k_F \gg 1$ and an intermediate regime with $m/k_F \sim 1$. When $m/k_F \to 0$ we recover the CFT result, in agreement again with our general argument above that the eigenvalues of density matrix are independent of $k_F$ in this case. However, when $m/k_F \neq 0$, the numerical simulation exhibits oscillating terms with period $\pi$ as function of $k_FL$. A fit (see below) verifies the analytic prediction (\ref{eq:Inprediction}).

\begin{figure}[htb!]
    \centering
    \includegraphics[width=0.96\linewidth]{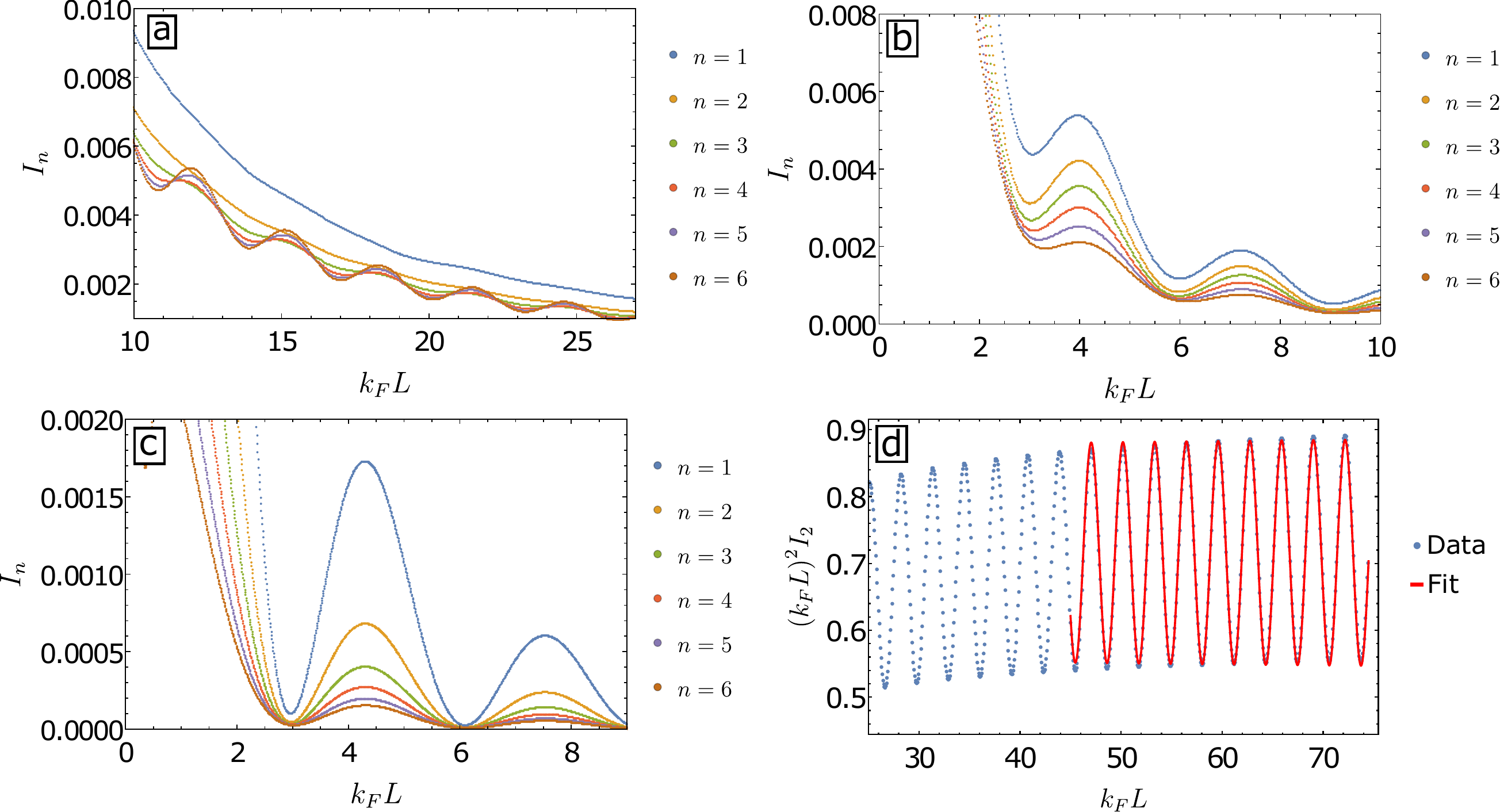}
    \caption{The subfigures a), b) and c) show simulations of $I_n(k_FL)$ for fixed $r$ for massive Dirac fermions. The parameters are $(k_Fr, m ,\frac{m}{k_F}) = (2, \frac{1}{500}, 0.2);\,(0.5, \frac{1}{200}, 1);\,(0.2, \frac{1}{100}, 5)$, respectively. We observe Friedel oscillations of  period $\pi$ as a function of $k_FL$, with amplitude dependent on $\frac{m}{k_F}$. In the subfigure d) we check the agreement with (\ref{eq:Inprediction}). We fit $(k_FL)^2I_2(k_FL)$ in the limit $k_FL \gg 1$ with $\frac{m}{k_F}=1$, $k_Fr=\frac{5}{3}$ and $k_F=\frac{1}{60}$. The expression used was $(k_FL)^2 I_2(k_FL)=A+B\cos(2k_FL)(k_FL)^C$; applying the fit to increasing ranges of $k_F L$ gives $C \to 0$, in agreement with (\ref{eq:Inprediction}).}
      \label{fig:In_all2}
 \end{figure}

Next, we present our results for the Renyi mutual information in Fig.~\ref{fig:In_all2}. Panels a), b) and c) show the behavior for different masses and different values of the Renyi parameter $n$. Finally, in panel d) we verify that the OPE limit prediction (\ref{eq:Inprediction}) is in excellent agreement with the numerical results. Let us stress again that these Friedel-type oscillations come from exchange of operators like (\ref{eq:exchange}) between the two regions; this effect is distinct from the boundary effects that determine the Friedel oscillations in the Renyi entropies. The numerical results are also consistent the cancellation of such contributions in the mutual information. 

\section{Measures of distinguishability}\label{sec:relrenyi}

In this last section of our analysis we turn our attention to a different question in quantum information: how to distinguish two different density matrices $\rho$ and $\sigma$. The main measure for this is the relative entropy; we will also study certain one-parameter generalizations of it. As discussed before, in QFT we can assign a density matrix to a region $V$ in space, obtained from the ground state by tracing over the Hilbert space of the complement. In the context of renormalization group flows, two states naturally arise: the one associated to the UV fixed point (call it $\sigma$), and another one ($\rho$) from the perturbed theory along the flow. A measure of distinguishability between $\sigma$ and $\rho$, like the relative entropy, provides a quantity that is monotonic under increase of the region $V$, and encodes nonperturbative information on the flow~\cite{Casini:2016fgb, Casini:2016udt, Casini:2017vbe, Lashkari:2017rcl, Casini_Renyi_2018}. In our setup of finite density field theory, we have an additional motivation for looking at these quantities, which is that we would like to measure the distinguishability between states in different superselection charge sectors. Let us present first the discussion on the relative entropy, and focus afterwards on its generalizations.

\subsection{Relative entropy}
\label{subsec:relative}

The relative entropy between two states $\rho$ and $\sigma$ is given by
\be
S_{rel}(\rho|\sigma)= \tr (\rho \log \rho) - \tr(\rho \log \sigma)\,.
\ee
In terms of the modular Hamiltonian of $\sigma=e^{-K_\sigma}/\tr(e^{-K_\sigma})$, it can be written as a difference of free energies,
\be
\label{eq:relative_delta}
S_{rel}(\rho|\sigma)= \Delta \langle K_\sigma \rangle - \Delta S
\ee
where $ \Delta \langle K_\sigma \rangle= \tr [(\rho - \sigma) K_\sigma]$ and $\Delta S= S(\rho) - S(\sigma)$. Operationally, it is a measure of distinguishability between the two states. It vanishes when the two states are the same, and becomes infinite if $\sigma$ is pure and different from $\rho$. An important property is that the relative entropy does not increase when restricted to a subsystem. In QFT, this means that the relative entropy increases when increasing the size of the region.

In general, it is not easy to calculate the relative entropy analytically between two states. However, it can be done when we compare a CFT to a CFT at finite density. Recall that both have the same density matrix eigenvalues, but the eigenfunctions are different, and the relative entropy should capture this.
Let $\sigma$ denote the reduced density matrix with $m= k_F =0$ (the conformal fixed point), and let $\rho$ be the density matrix for $m=0$ but $k_F \neq 0$.
The modular Hamiltonian for a CFT ground state reduced to a region $V$ that is a sphere of radius $R$ is known explicitly~\cite{Casini:2011kv},
\be\label{eq:Kcft}
K_\sigma= 2\pi\int_V d^{d-1}x \,\frac{R^2 - \vec x^2}{2R} T^{00}(x) + c'\,,
\ee
where the Cauchy surface is at constant time. It arises by conformally mapping the sphere to Rindler space, and noting that the modular Hamiltonian in the later case is just the boost generator.

In our case, $T^{00}$ is the energy-momentum tensor for a massless relativistic Dirac fermion,
\be
T^{00}=\frac{ i}{2} ( \bar \psi \gamma^0 \partial^0 \psi - ( \partial^0 \bar \psi) \gamma^0 \psi)\,,
\ee
so we can evaluate its expectation value in $\rho$ and $\sigma$ by taking coincidence limits of the appropriate Green's function in the finite charge and zero charge theories. Namely,
\be
\langle T^{00} \rangle =i \lim_{x \to y}\, \partial_{y_0} \langle \psi^\dag(x) \psi(y) \rangle\,.
\ee
The point-splitting divergence cancels out in $\Delta  \langle T^{00} \rangle$; at equal times we find
\be
\Delta  \langle T^{00} \rangle= \frac{ k_F^2}{2\pi}\,.
\ee
Replacing this into (\ref{eq:Kcft}) and performing the integral there for $d=2$ and an interval of length $r=2R$, we arrive at
\be
\label{eq:relative_entropy_m0}
S_{rel}(\rho|\sigma)= \Delta \langle K_\sigma \rangle = \frac{1}{6} k_F^2 r^2\,.
\ee

We thus find a super-extensive behavior in the relative entropy, coming from the modular Hamiltonian. This is a measure of the distinguishability between the massless fermionic ground state with zero charge and finite charge. We will find the same behavior below in numerical results. Interpreting the relative entropy as a difference of free energies as in (\ref{eq:relative_delta}), we see that changing to a different superselection charge sector is like a reversible adiabatic process: there is change in energy but the entropy stays constant. Here, this is a consequence of the fact that both states have the same density matrix eigenvalues, and hence $\Delta S=0$.

\subsection{Relative Renyi entropies}

There exists an interesting one-parameter generalization of relative entropy~\cite{muller2013quantum,wilde2014strong}
\be
S_\alpha(\rho|\sigma)=-\frac{1}{1-\alpha}\,\log\,\Tr\,\left(\sigma^{\frac{1-\alpha}{2\alpha}}\rho \sigma^{\frac{1-\alpha}{2\alpha}} \right)^\alpha\, ,
\label{eq:Salpha}
\ee
for $\alpha \in (0,1) \cup (1, \infty)$ and
\begin{equation}
\begin{split}
S_1(\rho| \sigma) &= \Tr\big( \rho (\log \rho -\log \sigma) \big), \\
S_\infty (\rho| \sigma) &= \log \parallel \sigma^{-1/2} \rho \sigma^{-1/2}  \parallel_\infty.
\end{split}
\end{equation}
These are usually referred to as relative Renyi entropies.

Let us focus on the range $\frac{1}{2}\le\alpha \le 1$. When $\alpha=1/2$,  (\ref{eq:Salpha}) gives the fidelity distance,
\bea
S_{1/2}(\rho|\sigma)&=&-2 \,\log\,\Tr\,\sqrt{\sigma^{1/2}\rho \sigma^{1/2}}=-2\,\log F(\rho,\sigma) \,,
\label{fid1}
\eea
where  $F(\rho,\sigma)$ denotes the quantum fidelity. Therefore, the measures for $\frac{1}{2}\le\alpha \le 1$ interpolate between quantum fidelity and quantum relative entropy.

The $S_\alpha$ have various nice properties. They are monotonically increasing in $\alpha$~\cite{muller2013quantum,beigi2013sandwiched, doi:10.1063/1.4838835}
\be\label{eq:alphamon}
\frac{d}{d\alpha}S_\alpha(\rho|\sigma)\geq 0 \,.
\ee
Since both the fidelity distance and the relative entropy are positive, and equal to zero only when $\rho= \sigma$, the same properties hold for the $S_\alpha$,
\be\label{eq:posit}
S_\alpha(\rho|\sigma) \ge 0\;\;,\;\; S_\alpha(\rho|\sigma) = 0 \;\;\text{for}\;\;\rho= \sigma\,.
\ee
Another important  property is monotonicity when increasing the size of the algebra. For two regions $V \subset \tilde V$, then
\be\label{eq:monoton}
S_\alpha(\rho_V |\sigma_V) \le S_\alpha(\rho_{\tilde V} |\sigma_{\tilde V}), \,
\ee
for $\alpha \ge 1/2$. They also admit a representation similar to Uhlmann's theorem for the fidelity~\cite{uhlmann1976transition} as maximizing over purifications; see~\cite{Casini_Renyi_2018} for a review of this.

These properties make the $S_\alpha$ interesting quantum information measures for QFT.
We will compute these quantities for Dirac fermions at finite density, by comparing two states $\sigma$ and $\rho$ corresponding to different choices of $m$ and $k_F$. 
For free fermions, an explicit expression for the relative Renyi entropies can be found in terms of the two point correlation function (see e.g.~\cite{Casini_Renyi_2018}),
\begin{equation}
\label{eq:QRRE_T}
S_\alpha (\rho | \sigma) = -\frac{1}{1-\alpha}\log \frac{\det \big[1 + \big( T^{ \frac{1-\alpha}{2 \alpha} }T'T^{ \frac{1-\alpha}{2 \alpha} } \big)^\alpha]^{\frac{1}{2}}}{\big(\det [1+T] \big)^{\frac{1-\alpha}{2}} \big(\det [1+T'] \big)^{\frac{\alpha}{2}}},
\end{equation}
where 
\be
\label{eq:Tmat_def}
T=\frac{1+\mathcal{C}}{1-\mathcal{C}}, \quad T^T = T^{-1}, \quad T^\dagger = T\, ,
\ee
and $\mathcal{C}_{IJ} = \frac{1}{2}\langle [w_I, w_J] \rangle$ is defined in terms of the Majorana operators $w_I = (\psi_j + \psi^\dagger_j,i(\psi_j - \psi^\dagger_j) )$. The correlation matrix $\mathcal{C}$ can be written in terms of $C_{ij} = \langle \psi^\dagger_i \psi_j \rangle$,
\be
\label{eq:mathcalC_to_C}
\mathcal{C} = \begin{pmatrix}
2 i \Im C^T &  -i (1 - 2 \Re C^T) \\
i (1 - 2 \Re C^T) & 2 i \Im C^T
\end{pmatrix}.
\ee

\subsection{Numerical results}

We proceed to evaluate numerically (\ref{eq:QRRE_T}) for different states parametrized by the values of the two relevant couplings $m, k_F$. We also take $\alpha \to 1$ in order to obtain the relative entropy.\footnote{Again, we have to divide the numerical result by two in order to account for the fermion doubling.}

\begin{figure}[h]
\centering
\includegraphics[width=0.8\linewidth]{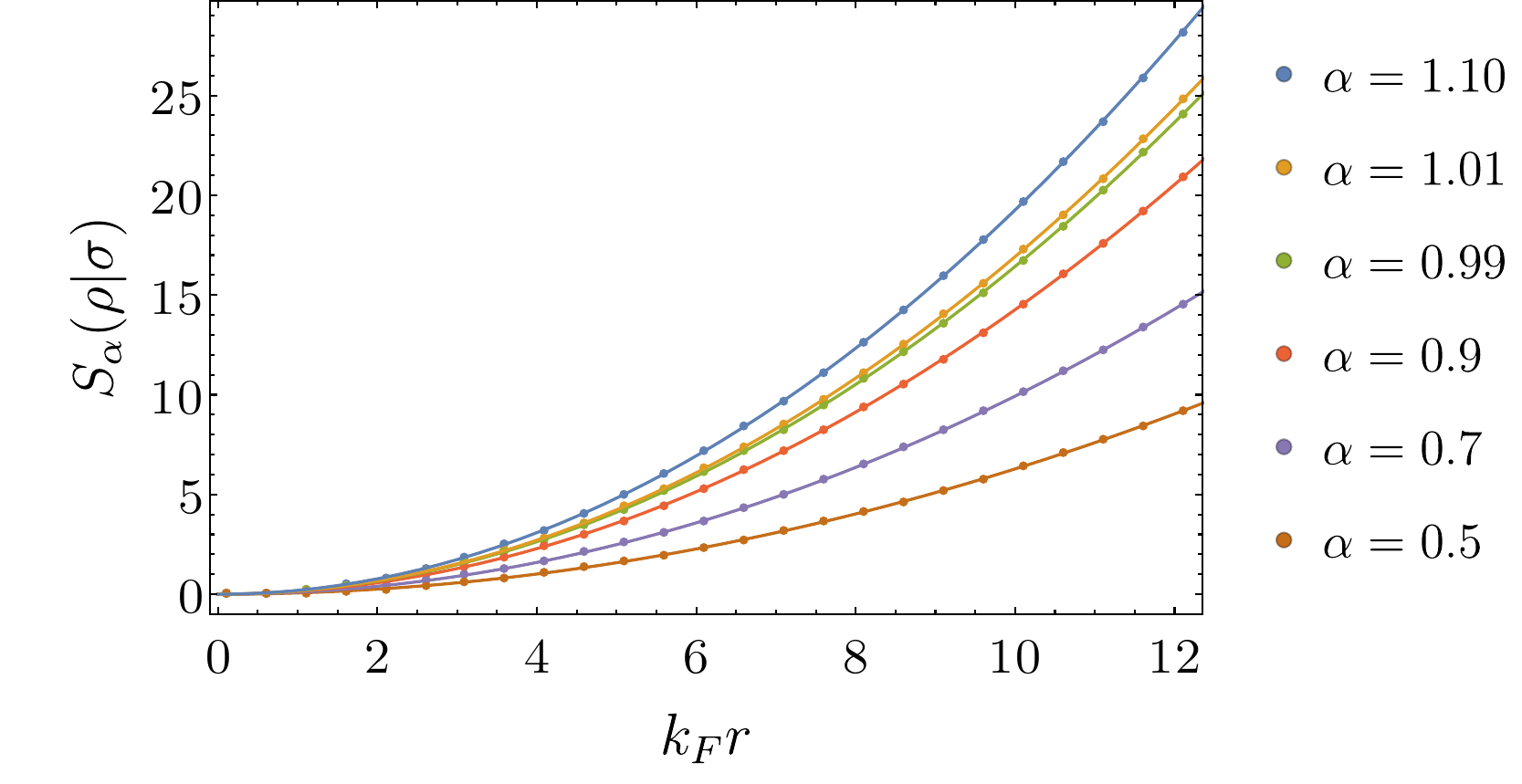}
\caption{Plot of $S_\alpha$ with $\alpha \in [0.5, 1.1]$ for $m=0$, comparing states with $k_F=0$ and $k_F =1/20$. A functional dependence $S_\alpha=A_{\alpha} (k_F r)^2$ is observed and plotted for all $\alpha$ considered. When $\alpha \to 1$ the plots create a bound from above and from below to the relative entropy. The coefficients of the fits are approximately $A_{\alpha}=0.164$ for $\alpha=0.99$ and $A_{\alpha}=0.169$ for $\alpha=1.01$; this agrees with the analytic prediction $A_1=1/6$ for $\alpha=1$ in (\ref{eq:relative_entropy_m0}).}
\label{fig:Salpha_constant_m}
\end{figure}

Let us begin by discussing the sates with fixed $m=0$, comparing $k_F = 0$ ($\sigma$) and $k_F = 1/20$ ($\rho$). The results are shown in Fig.~\ref{fig:Salpha_constant_m}.

We find that all curves follow the functional dependence $S_\alpha =A_{\alpha}(k_F r)^2$, for a constant parameter $A_{\alpha}$. 
This means that the Renyi relative entropies exhibit a super-extensive dependence on $k_F r$. For $\alpha \to 1$, this agrees with the analytic results for the relative entropy in Section~\ref{subsec:relative}; in particular, the value $A_1=\frac{1}{6}$ found in~(\ref{eq:relative_entropy_m0})
 is close to the lower and upper bounds given by the numerical results with $\alpha \sim 1$. In this case, the super-extensive dependence follows directly from the form (\ref{eq:Kcft}) of the modular Hamiltonian for a CFT. At present we do not have a similar analytic understanding for the nonlinear relative entropies with $\alpha \in [1/2,1)$, and it would be interesting to revisit this question in future work.

\begin{figure}[h]
\centering
\includegraphics[width=0.8\linewidth]{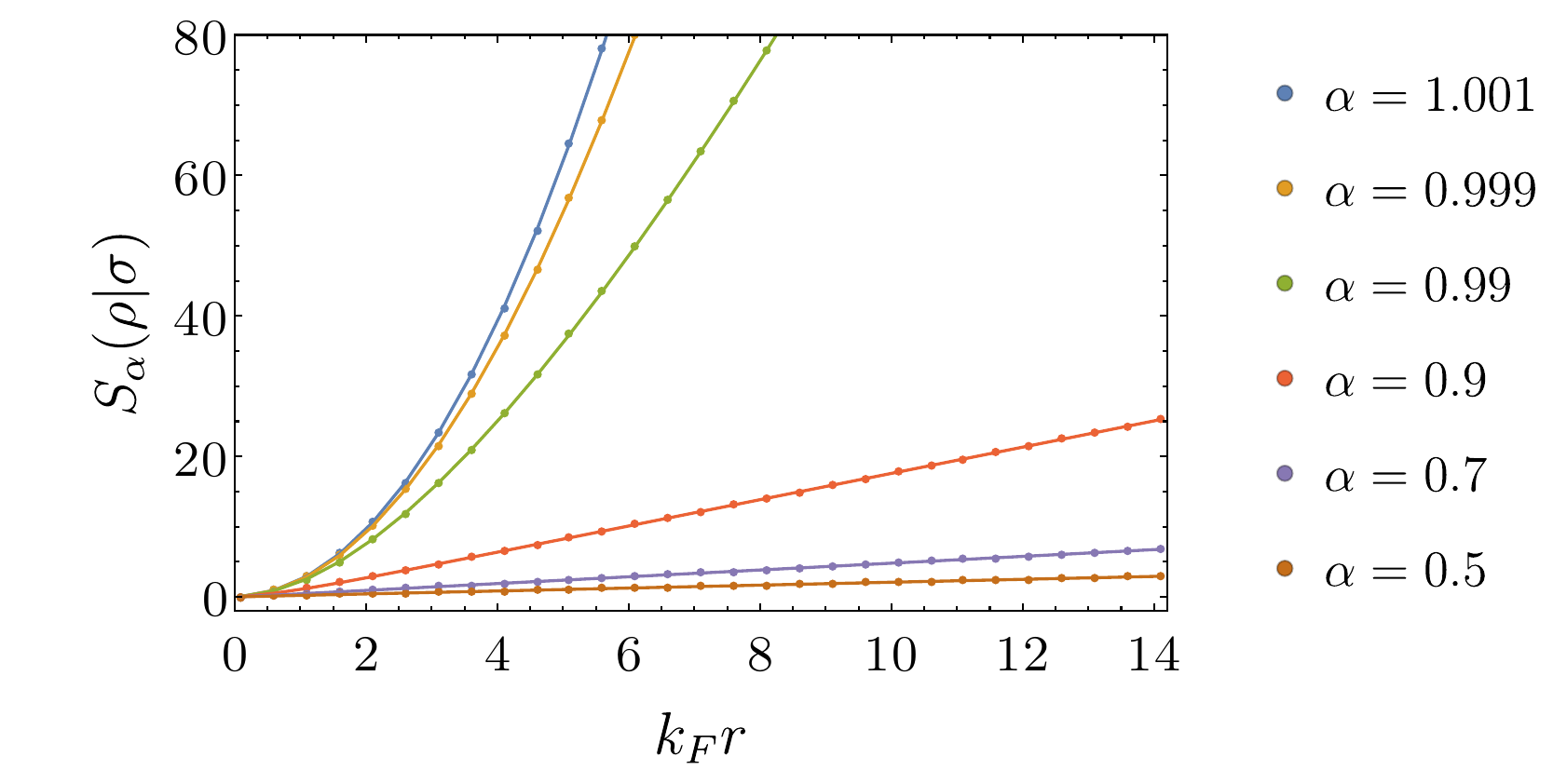}
\caption{Plot for the relative Renyi entropies with $\alpha \in [0.5, 1.1]$ for $k_F=1/20$ and comparing states with $m=0$ and $m = 1/10$. The curves scale like  $\sim (k_F r)^2$ for small $k_F r$, but higher powers also arise for larger values.}
\label{fig:Salpha_constant_kF}
\end{figure}

Finally, let us discuss the case of two states with the same $k_F$ and different masses; 
the reference state $\sigma$ corresponds to the theory with $m =0$, while $\rho$ has $m \neq 0$. Both states are in the same charge sector, and the $S_\alpha$ measure the distinguishability associated to the RG flow caused by the mass. In particular, at low energies both theories have different Fermi velocities.
Numerical results for this situation are shown in Fig.~\ref{fig:Salpha_constant_kF}. 

Unlike the previous case, it is now harder to develop an analytic understanding. We can already see this for the relative entropy, $\alpha=1$, for which the modular Hamiltonian is not known explicitly.\footnote{Other recent lattice works on the modular Hamiltonian include~\cite{Eisler:2019rnr, Eisler:2020lyn}.} The numerical results exhibit a dependence $S_\alpha \sim (k_F r)^2$ when $k_Fr \ll 1$, which is modified at larger values of $k_F r$. It is worth emphasizing that these quantities are always monotonic, and as such they could provide nontrivial information about the RG flow even for nonrelativistic QFT. In this direction, it would be interesting to relate their behavior more explicitly to the dynamics of the theory. In particular, the monotonicity ensures the absence of Friedel-type oscillations such as those detected in previous sections. It would be interesting to pursue this direction further in future work.

\section{Conclusions and future directions}\label{sec:concl}
In this article we studied several measures of quantum information in finite density field theory. We focused on free massive Dirac fermions at finite density in $1+1$ space-time dimensions. Despite the simplicity of the setup, some of the lessons reflected in these measures are likely more general, and we would like to emphasize them here, together with future directions suggested by our results.

We established that the entropic $c$-function (constructed from the logarithmic derivative of the entanglement entropy) is not monotonic. This is in stark contrast with the behavior in Lorentz-invariant QFT, where this quantity obeys $c'(r) \le 0$, and this leads to the C-theorem~\cite{Zamolodchikov:1986gt, Casini:2004bw}. We found $c_{UV}=c_{IR}$, but with nontrivial intermediate behavior that encoded the competition between the charge density and the mass gap. Curiously, since the UV and IR central charges are the same, the weak version of the theorem ($c_{UV} \ge c_{IR}$) is not violated. However, the relativistic sum rule implies that if $c_{UV}=c_{IR}$ then there is no RG flow (see e.g.~\cite{Cappelli:1990yc}), and this is explicitly violated in our setup. In any case, the entropic $c$-function provides a finite quantity that is sensitive to the creation of entanglement associated to the Fermi surface, and it would be interesting to study its behavior in interacting systems numerically. 

The analysis of Renyi entropies also revealed nontrivial effects due to the Fermi surface, most notably Friedel-type oscillations that modify the CFT result at subleading order. These were observed before in lattice models~\cite{Calabrese2010ParityEI}; we argued that they also arise in the continuum theory when the chiral symmetry is absent (broken by the mass term).
Using a defect operator product expansion, these oscillations are found to arise from operators of fractional dimension localized in the conical singularities of the replicated manifold, in agreement with previous suggestions~\cite{Swingle_2013, Cardy2010UnusualCT}. The same argument applies to interacting models, so this oscillatory dependence could be interesting for probing non-Fermi liquids. 
 
The mutual information is an interesting and promising observable for theories of quantum matter, because it measures correlations between different regions.
We proved that, in a long distance expansion, the mutual information detects the Fermi surface already at leading order via new oscillatory terms. These are distinct from the Friedel oscillations in the Renyi entropies. We also presented a generalization that accounts for nontrivial anomalous dimensions. These features suggest that the mutual information and its Renyi versions are very promising probes for correlated systems.  It would also be interesting to allow for time-dependent perturbations and include effects from out-of-time ordered correlators. See e.g.~\cite{Islam:2015, Li:2017} for recent experimental progress on these fronts.

Finally, we studied the relative entropy (and its uni-parametric generalizations) as a measure of distinguishability between different quantum states. The relative entropy exhibits a super-extensive behavior for states in different superselection charge sectors, and it is monotonic and finite. These properties make it useful for understanding nonperturbative aspects of the RG at finite density. To continue along this line, it would be important to determine how to extract intrinsic properties of fixed points from the relative entropy. This was done for relativistic theories in~\cite{Casini:2016fgb, Casini:2016udt}. The structure of the energy-momentum tensor correlators at finite density could also provide complementary information.

It would be important to extend the present study to $d>2$. In this case, Fermi surfaces feature a logarithmic violation of the entropic area law~\cite{PhysRevLett.96.010404}, while the leading area term in the EE is power-law divergent. This is related to hyperscaling violation~\cite{Fisher:1986zz, Huijse:2011ef, Dong:2012se}, which is absent in $1+1$ dimensions. A fruitful direction may be to try to construct finite entropic quantities from the EE or the relative entropy. It would of course be extremely interesting to also consider interacting theories; as we have discussed before, some of our methods extend to the interacting case. Along this line, it would be nice to evaluate the quantum information measures explored here in holographic models, with the goal of shedding more light on the elusive holographic Fermi surfaces~\cite{Hartnoll:2016apf}.

\section*{Acknowledgments}
We thank H.~Casini for many interesting discussions and comments on the manuscript. LD is supported by CNEA and UNCuyo, Inst. Balseiro. RM is supported by IST Austria. MS is supported by CONICET and UNCuyo, Inst. Balseiro.
GT is supported by CONICET (PIP grant 11220150100299), ANPCyT (PICT 2018-2517), CNEA, and UNCuyo, Inst. Balseiro.

\appendix

\section{Fermions on the lattice}
\label{app:lattice}

In this Appendix we collect some results on lattice fermions that are used in the main text.

\subsection{Eigenvectors and correlator}

To compute $C_{ij}$ we choose the chiral basis $\gamma^0=\sigma^1$, $\gamma^1=i\sigma^2$ and therefore $\gamma^3=\gamma^0\gamma^1=-\sigma^3$. The eigenvectors are
\begin{equation}
v_+(k)=N_+\begin{pmatrix}
m \\ \sin(k)+\sqrt{m^2+\sin(k)^2}
\end{pmatrix}\:\:\:\: \text{y}  \:\:\:\: v_-(k)= N_-\begin{pmatrix}
m \\ \sin(k)-\sqrt{m^2+\sin(k)^2}
\end{pmatrix},
\end{equation} 
with normalization constants $N_{\pm}$ as a consequence of the unitarity of $U(k)=\left(v_+(k) \;,\; v_-(k)\right)$
\begin{equation}
\begin{split}
N_+^2 & =\frac{1}{2\sqrt{m^2+\sin(k)^2}(\sqrt{m^2+\sin(k)^2}+\sin(k))}, \\ N_-^2 & =\frac{1}{2\sqrt{m^2+\sin(k)^2}(\sqrt{m^2+\sin(k)^2}-\sin(k))}\,.
\end{split}
\end{equation}

Using $m^2=\left(\sqrt{m^2+\sin(k)^2}-\sin(k)\right)\left(\sqrt{m^2+\sin(k)^2}+\sin(k)\right)$,
the outer product of the eigenvectors becomes
\begin{equation}\label{eq:vp}
v^{\dag}_+(k)v_+(k)=\begin{pmatrix}
\frac{1}{2}-\frac{\sin(k)}{2\sqrt{m^2+\sin(k)^2}} & \frac{m}{2\sqrt{m^2+\sin(k)^2}} \\ \frac{m}{2\sqrt{m^2+\sin(k)^2}} & \frac{1}{2}+\frac{\sin(k)}{2\sqrt{m^2+\sin(k)^2}}
\end{pmatrix} = \frac{1}{2}\mathbb{I}+\frac{m\gamma^0}{2\sqrt{m^2+\sin(k)^2}}+\frac{\sin(k)\gamma^0\gamma^1}{2\sqrt{m^2+\sin(k)^2}},
\end{equation}
\begin{equation}\label{eq:vm}
v^{\dag}_-(k)v_-(k)=\begin{pmatrix}
\frac{1}{2}+\frac{\sin(k)}{2\sqrt{m^2+\sin(k)^2}} & -\frac{m}{2\sqrt{m^2+\sin(k)^2}} \\ -\frac{m}{2\sqrt{m^2+\sin(k)^2}} & \frac{1}{2}-\frac{\sin(k)}{2\sqrt{m^2+\sin(k)^2}}
\end{pmatrix} = \frac{1}{2}\mathbb{I}-\frac{m\gamma^0}{2\sqrt{m^2+\sin(k)^2}}-\frac{\sin(k)\gamma^0\gamma^1}{2\sqrt{m^2+\sin(k)^2}}.
\end{equation}

\subsection{Fermion doubling at finite density}
\label{app:doubling}

The Dirac field on the lattice has been studied intensively in the past, starting from~\cite{CHODOS1977426, CHODOS1977second,Susskind1976}. These works noticed the fermion doubling problem on the lattice; we need to revisit this issue because of some new effects at finite density that need to be taken into account.

As noticed in these works, the dispersion relation (\ref{eq:eppm}) is symmetric under $k \to k+\pi$. This is a direct consequence of a discrete symmetry in the Hamiltonian (\ref{eq:Hlattice}), 
\begin{equation}
\psi_n \to S \psi_n\;,\;\; S=(-1)^n \gamma^0\,.
\end{equation}
In the continuum low energy limit, we keep modes with $k \sim 0$ but eliminate those modes with high momentum $k \sim \pi$. This is accomplished in terms of the combination
\begin{equation}
\Psi(x)\equiv\lim_{a\to 0}\frac{\psi_{n+1} + \psi_n}{2} ,
\label{eq:averag_doubling} 
\end{equation}
for $x=na$. This is an eigenvector of the discrete symmetry, $S\psi_n=\psi_n$.

\begin{figure}[h]
\centering
\includegraphics[width=1\linewidth]{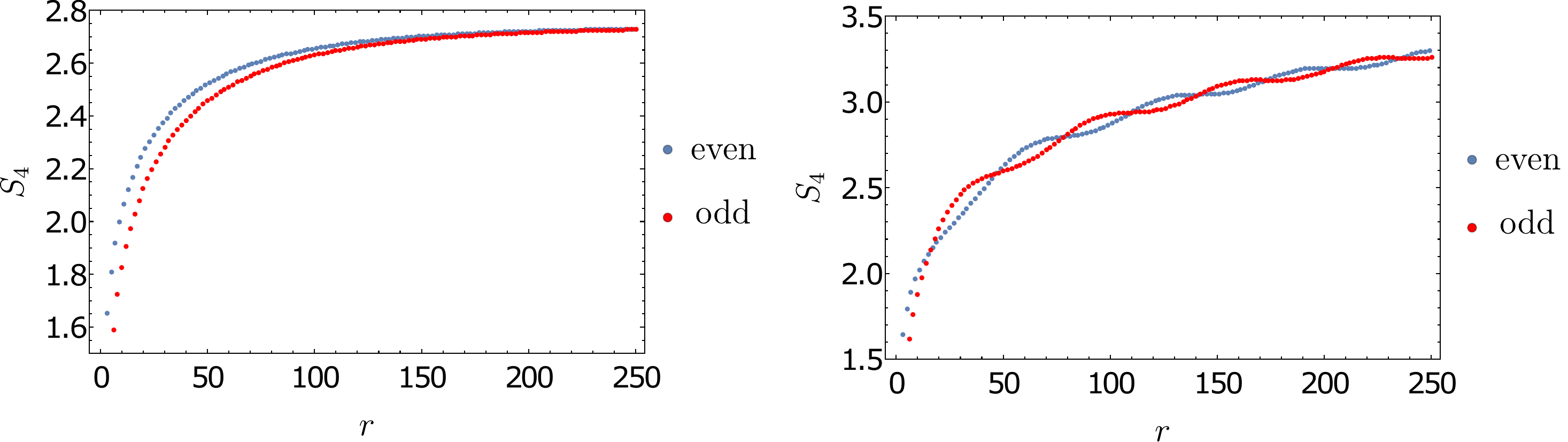}
\caption{Left panel: $S_4$ for $m=1/200$ and $k_F=0$, distinguishing the parity $(-1)^n$ of $r$. Both branches coincide when $r\gg 1$. Right panel: $S_4$ for $m=1/200$ and $k_F=1/20$; the even and odd branches are shifted by $\pi/2k_F$ as a function of $r$.}
\label{fig:doubling}
\end{figure}

In our context, when $k_F=0$ this doubling is simply taken into account by dividing the entropies by two; see the left picture on Fig.~\ref{fig:doubling}. However, when $k_F \neq 0$ we cannot just do that because we find that the two branches exhibit oscillatory behavior and are not in phase; this is illustrated in the right panel of Fig.~\ref{fig:doubling}.  More precisely, the curves for even and odd sites are shifted by $\frac{\pi}{2}$ as a function of $k_Fr$. Averaging the two curves would not give the right result for the Renyi entropies. The origin of this shift can be seen already when $m=0$ and $k_F \neq 0$, from the form of the correlator
\begin{equation}
C_{ij}=C_{ij}^0+\begin{cases}
\mathbb{I}\frac{\sin[k_F(i-j)]}{\pi(i-j)}\;\; i-j \;\;\text{even}\\ i \gamma^0 \gamma^1\frac{\cos[k_F(i-j)]-1}{\pi(i-j)} \;\; i-j \;\;\text{odd},
\end{cases}
\end{equation}
with $k_F=\arcsin\left(\sqrt{\mu_F^2-m^2}\right)$ and
\begin{equation}
C_{ij}^0=\mathbb{I}\frac{\delta_{ij}}{2}+i\gamma^0\gamma^1\frac{1-(-1)^{(i-j)}}{2\pi (i-j)}.
\end{equation}
Both branches are shifted by $k_F(i-j)\to k_F(i-j) \pm \frac{\pi}{2}$ in the finite density coefficients, which becomes $k_Fr \to k_Fr \pm \frac{\pi}{2}$ in the continuum limit. When $m \neq 0$ we do not have an analytic expression, but a similar behavior is seen in numerical calculations. We stress that, unlike the fermion doubling, this shift is a property of the continuum. To address this issue, we have chosen data given by even sites, and furthermore divided the Renyi entropies and other measures by two to take into account the doubling.

\bibliography{EE}{}
\bibliographystyle{utphys}

\end{document}